\documentclass[prb,twocolumn,showpacs,preprintnumbers,amsmath,aps]{revtex4}
\usepackage{graphicx,hyperref}% Include figure files
\usepackage{amsmath}
\usepackage{bm}
\usepackage{subfigure}% bold math

\newcommand\vect[1]{ \boldsymbol{ #1}}
\def\nbC{{\mathchoice {\setbox0=\hbox{$\displaystyle\rm C$}%
\hbox{\hbox to0pt{\kern0.4\wd0\vrule height0.9\ht0\hss}\box0}}
{\setbox0=\hbox{$\textstyle\rm C$}\hbox{\hbox
to0pt{\kern0.4\wd0\vrule height0.9\ht0\hss}\box0}}
{\setbox0=\hbox{$\scriptstyle\rm C$}\hbox{\hbox
to0pt{\kern0.4\wd0\vrule height0.9\ht0\hss}\box0}}
{\setbox0=\hbox{$\scriptscriptstyle\rm C$}\hbox{\hbox
to0pt{\kern0.4\wd0\vrule height0.9\ht0\hss}\box0}}}}
\def\nbQ{{\mathchoice {\setbox0=\hbox{$\displaystyle\rm
Q$}\hbox{\raise
0.15\ht0\hbox to0pt{\kern0.4\wd0\vrule height0.8\ht0\hss}\box0}}
{\setbox0=\hbox{$\textstyle\rm Q$}\hbox{\raise
0.15\ht0\hbox to0pt{\kern0.4\wd0\vrule height0.8\ht0\hss}\box0}}
{\setbox0=\hbox{$\scriptstyle\rm Q$}\hbox{\raise
0.15\ht0\hbox to0pt{\kern0.4\wd0\vrule height0.7\ht0\hss}\box0}}
{\setbox0=\hbox{$\scriptscriptstyle\rm Q$}\hbox{\raise
0.15\ht0\hbox to0pt{\kern0.4\wd0\vrule height0.7\ht0\hss}\box0}}}}
\def\nbT{{\mathchoice {\setbox0=\hbox{$\displaystyle\rm
T$}\hbox{\hbox to0pt{\kern0.3\wd0\vrule height0.9\ht0\hss}\box0}}
{\setbox0=\hbox{$\textstyle\rm T$}\hbox{\hbox
to0pt{\kern0.3\wd0\vrule height0.9\ht0\hss}\box0}}
{\setbox0=\hbox{$\scriptstyle\rm T$}\hbox{\hbox
to0pt{\kern0.3\wd0\vrule height0.9\ht0\hss}\box0}}
{\setbox0=\hbox{$\scriptscriptstyle\rm T$}\hbox{\hbox
to0pt{\kern0.3\wd0\vrule height0.9\ht0\hss}\box0}}}}
\def\nbS{{\mathchoice
{\setbox0=\hbox{$\displaystyle     \rm S$}\hbox{\raise0.5\ht0%
\hbox to0pt{\kern0.35\wd0\vrule height0.45\ht0\hss}\hbox
to0pt{\kern0.55\wd0\vrule height0.5\ht0\hss}\box0}}
{\setbox0=\hbox{$\textstyle        \rm S$}\hbox{\raise0.5\ht0%
\hbox to0pt{\kern0.35\wd0\vrule height0.45\ht0\hss}\hbox
to0pt{\kern0.55\wd0\vrule height0.5\ht0\hss}\box0}}
{\setbox0=\hbox{$\scriptstyle      \rm S$}\hbox{\raise0.5\ht0%
\hboxto0pt{\kern0.35\wd0\vrule height0.45\ht0\hss}\raise0.05\ht0%
\hbox to0pt{\kern0.5\wd0\vrule height0.45\ht0\hss}\box0}}
{\setbox0=\hbox{$\scriptscriptstyle\rm S$}\hbox{\raise0.5\ht0%
\hboxto0pt{\kern0.4\wd0\vrule height0.45\ht0\hss}\raise0.05\ht0%
\hbox to0pt{\kern0.55\wd0\vrule height0.45\ht0\hss}\box0}}}}
\def\nbZ{{\mathchoice {\hbox{$\sf\textstyle Z\kern-0.4em Z$}}
{\hbox{$\sf\textstyle Z\kern-0.4em Z$}}
{\hbox{$\sf\scriptstyle Z\kern-0.3em Z$}}
{\hbox{$\sf\scriptscriptstyle Z\kern-0.2em Z$}}}}
\begin{document}

\title{Random-field Ising and $O(N)$ models: Theoretical description through the functional renormalization group}

\author{Gilles Tarjus} \email{tarjus@lptmc.jussieu.fr}
\affiliation{LPTMC, CNRS-UMR 7600, Sorbonne Universit\'e, 4 Pl. Jussieu, 75252 Paris c\'edex 05, France}

\author{Matthieu Tissier} \email{tissier@lptmc.jussieu.fr}
\affiliation{LPTMC, CNRS-UMR 7600, Sorbonne Universit\'e, 4 Pl. Jussieu, 75252 Paris c\'edex 05, France}

\date{\today}

\begin{abstract}
We review the theoretical description of the random field Ising and $O(N)$ models obtained from the functional renormalization group, either in its nonperturbative implementation or, in some limits, in perturbative implementations. The approach solves some of the questions concerning the critical behavior of random-field systems that have stayed pending for many years: What is the mechanism for the breakdown of dimensional reduction and the breaking of the underlying supersymmetry below $d=6$? Can one provide a theoretical computation of the critical exponents, including the exponent $\psi$ characterizing the activated dynamic scaling? Is it possible to theoretically describe collective phenomena such as avalanches and droplets? Is the critical scaling described by 2 or 3 independent exponents? What is the phase behavior of the random-field $O(N)$ model in the whole ($N$, $d$) plane and what is the lower critical dimension of quasi-long range order for $N=2$? Are the equilibrium and out-of-equilibrium critical points of the RFIM in the same universality class?
\end{abstract}


\maketitle

\section{Introduction}
\label{sec:intro}

The random-field Ising and $O(N)$ models are archetypal systems for describing the competition between an ordering tendency generated by interactions and a disordering one associated with the presence of a quenched disorder that directly couples to the local order parameter. These models provide a playground to investigate the consequences of such a competition on the collective behavior at large scale. In the simplest formulation, the models are described by a Hamiltonian
\begin{equation}
\mathcal H=- \frac 12 \sum_{i,j} J_{ij}\, \bm S_i \mathbf{\cdot} \bm S_j +\sum_i \bm h_i  \mathbf{\cdot}\bm S_i
\label{eq_latticeRFO(N)M}
\end{equation}
where $\bm S_i$ is an $N$-component (classical) spin on the vertex $i$ of a $d$-dimensional Euclidean lattice, $J_{ij}>0$ is a short-ranged ferromagnetic interaction, e.g., with $i$ and $j$ nearest-neighbor sites on the lattice, and $\bm h_i$ is a random field which is usually chosen for simplicity independently on each lattice site and is sampled from a given probability distribution with zero mean, $\overline{h_i^\mu}=0$, and finite variance, $\overline{h_i^\mu h_j^\nu}=\delta_{ij}\delta_{\mu \nu}\Delta_B$ with $\mu,\nu=1,\cdots, N$. The most common cases correspond $N=1$ (Ising), $N=2$ ($XY$), and $N=3$ (Heisenberg).

Although the random-field $O(N)$ models (RF$O(N)$M) are commonly formulated in the language of ferromagnetic systems (as above), it turns out that generating magnetic fields that are random on short length scales is far from straightforward in actual materials. It is only recently that this has been achieved in anisotropic dipolar magnetic insulators, which represent a realization of the random-field Ising model (RFIM) \cite{ferro_RFIM,ferro_RFIM-exp}. Otherwise, random-field models emerge as the effective theory for a host of systems in the presence of quenched disorder. In physics the experimentally most studied systems, which has been argued to be in the universality class of the RFIM \cite{diluteAFM} are diluted antiferromagnets in a uniform external field \cite{belanger_review}. Other examples include critical fluids in disordered porous media such as silica gels \cite{degennes_RFIM,chan-cannell,pitard_RFIM,vink_RFIM} for the $N=1$ version, vortex phases in type-II superconductors (elastic glass model) for the $N=2$ version \cite{giamarchi_bragg,blatter_review,giamarchi_bragg-review}, impurities in an incommensurate charge density wave in a tetragonal crystal, which describes vestigial nematicity in the pseudo-gap phase of the cuprates ($N=2$ and $N=1$) \cite{nie_PNAS,nematic2}, or the Mott metal-insulator transition in vanadium dioxide \cite{mott}. In addition, the RFIM has recently appeared as an effective description in the context of the glass transition of liquids \cite{wolynes_RFIM,franz_RFIM, giulio_RFIM}. 

The RFIM is also one of the simplest statistical-mechanical models that captures the anomalous irreversible collective response seen in a wide range of physical, biological, or socio-economic situations in the presence of attractive interactions and intrinsic heterogeneity or disorder \cite{sethna01}. When slowly driven at zero temperature, it displays as a function of disorder strength an out-of-equilibrium phase transition characterized by critical scaling and scale-free avalanches  (``crackling noise") \cite{sethna93,dahmen96,perkovic,perez04}. This description applies, for instance, to the Barkhausen noise observed in magnetic materials \cite{bertotti,sethna05} and in martensites \cite{planes-vives}, to the hysteresis behavior found in the fluid adsorption in a disordered porous solid \cite{rosinberg-monson,detcheverry04,wolf_aerogel},  the functioning of isometrically activated muscles \cite{truski_muscle}, the yielding transition of quasi-statically sheared amorphous solids \cite{ozawa_PNAS}, or to agent-based models in socio-economic context \cite{bouchaud13}.

The purpose of this article is to provide a short review of the theoretical description of random-field systems that has been obtained through the use of the functional renormalization group (FRG), whether in its nonperturbative or its perturbative implementations.

The paper is organized as follows. In Sec. \ref{sec_models} we present the models describing random-field systems with Ising and $O(N)$ symmetries as well as the physical situations to be described. The next section is devoted to a brief recap of the results prior to 2004 (which is when our first results using the FRG appeared \cite{tarjus04}) and is concluded by a (nonexhaustive) list of then-pending questions. In Sec. \ref{sec_avalanches} we discuss the collective events known as avalanches and droplets that are present in random-field systems and their consequences on correlation functions and on cumulants of the renormalized disorder. We stress the need for a multi-copy or multi-replica formalism in which the replicas have the same disorder but are coupled to distinct, independent, sources. In the following section we summarize the main results obtained by means of the FRG, with a focus on the long-distance equilibrium properties. The framework of the FRG for the random-field Ising and $O(N)$ model is described in Sec. \ref{sec_NPFRG}. We sequentially sketch the exact FRG approach and the derivation of exact functional flow equations for the cumulants of the renormalized disorder (\ref{sec_NPFRG}-A,B), the nonperturbative approximation scheme (\ref{sec_NPFRG}-C), and the final (functional) fixed-point equations and their solution (\ref{sec_NPFRG}-D). This is then followed in Sec. \ref{sec_robustness} by a discussion of the robustness of the nonperturbative FRG results and a presentation of perturbative but functional RG approaches in two limiting cases: near the lower critical dimension for long-range ferromagnetism, $d=4$, for the RF$O(N>1)$M and near the upper critical dimension, $d=6$, for the RFIM. Sec. \ref{sec_further} is a short account of additional results obtained through the FRG, and we conclude in Sec. \ref{sec_conclusion}.

\section{Models}
\label{sec_models}

Since we are interested in the long-distance and long-time physics of random-field systems, it is convenient to start with the field-theoretical version of Eq. (\ref{eq_latticeRFO(N)M}). We then consider the following ``bare action''  for an $N$-component scalar field $\bm\varphi$ in $d$-dimensional space,
\begin{equation}
\begin{aligned}
\label{eq_ham_dis_RFIM}
&S[\bm\varphi; \bm h]=  S_B[\bm\varphi]-\int_{x} \bm h_x \mathbf{\cdot} \bm\varphi_x \,, \\&
S_B[\bm\varphi]= \int_{x}\bigg\{\frac{1}{2}\vert \partial_x \bm \varphi_x\vert^2+ \frac{r}{2} \vert \bm\varphi_x\vert ^2 + \frac{u}{4!} \vert \bm\varphi_x\vert^4 \bigg\},
\end{aligned}
\end{equation}
where $ \int_{x} \equiv \int d^d x$ and $\bm h$ is a random ``source'' (a random magnetic field); this quenched random field $\bm h$ is sampled from a distribution characterized by a zero mean and  a variance $\overline{h_x^\mu h_y^\nu}= \Delta_B \delta_{\mu\nu}\delta^{(d)}( x-y)$. This model corresponds to systems with short-ranged interactions and short-ranged correlations of the random field. The extension to long-ranged interactions and/or disorder correlations will be discussed in Sec. \ref{sec_further}. An ultraviolet (UV) cutoff $\Lambda$ on the momenta, associated with the inverse of a microscopic length scale such as a lattice spacing, is also implicitly taken into account.

Models with quenched random fields can be, and have been, studied in different physical situations. First, they have been considered {\it in thermodynamic equilibrium}. The relevant quantity is then the sample-dependent partition function
\begin{equation}
\label{eq_partition_disorder}
\mathcal Z[\bm J; \bm h]=\int \mathcal D \bm\varphi \, {\rm{exp}}\big [-S[\bm\varphi; \bm h] +\int_x \bm J(x) \mathbf{\cdot}\bm\varphi(x)\big ]
\end{equation}
where $\bm J$ is an $N$-component external source (magnetic field). The thermodynamics of the system is described by the average over quenched disorder  of the free-energy functional, i.e., of the logarithm of the partition function, $\mathcal W[\bm J;\bm h]=\ln \mathcal Z[\bm J; \bm h]$. There is, however, more to the problem than this average free energy, and we will discuss in more detail below the difficulties associated with the fact that properties in a disordered system are {\it a priori} sample dependent. Note that when studying the equilibrium critical point that takes place when $\bm J=\bm 0$ because of the statistical $Z_2$ or $O(N)$ symmetry of the theory (for symmetric distributions of the random field), $\bm J(x)$ is just a standard tool to generate correlations functions by functional differentiation \cite{zinnjustin89}.

The models can also be investigated in equilibrium but directly {\it at zero temperature}, where one then focuses on the properties of the ground state. The latter is solution of the following stochastic field equation,
\begin{equation}
\label{eq_stochastic_GS}
\frac{\delta S_B[\bm\varphi]}{\delta \varphi^\mu_{x}}-h^\mu(x)-J^\mu_{x}=0\,,
\end{equation}
which is obtained by minimizing the action in Eq. (\ref{eq_partition_disorder}). The ground-state configuration $\bm \varphi_{GS}(x)$ corresponds to the solution with lowest energy (or action). It is in this context of the equilibrium properties at zero temperature that Parisi and Sourlas \cite{parisi79} have developed their supersymmetric construction, on which we will comment more below.

Finally, one may consider the {\it dynamics} of random-field systems, {\it either near to equilibrium or far from it}. At a coarse-grained level, this can be described by a Langevin equation,
\begin{equation}
\label{eq_stochastic_dynamics_RFIM}
\partial_t\varphi^\mu_{xt}=-\frac{\delta S_B[\varphi]}{\delta \varphi^\mu_{xt}}+h^\mu(x)+J^\mu_{xt} + \eta^\mu_{xt},
\end{equation}
where $\bm\eta_{xt}$ is a Gaussian random thermal noise with zero mean and variance $\langle\eta^\mu_{xt}\eta^\nu_{x't'}\rangle=2T \delta_{\mu\nu}\delta^{(d)}(x-x')\delta(t-t')$. The relaxation dynamics to equilibrium corresponds to taking $T>0$ and $\bm J$ independent of time. On the other hand, the situation in which the system is quasi-statically driven by a slowly varying applied source corresponds to $T=0$ and $\bm J_{xt}=\bm J+\bm\Omega t$, with $\vert\bm\Omega\vert \to 0^\pm$ depending on whether the source is increased or decreased \cite{footnote_bulk-vs-interface_dynamics}. This out-of-equilibrium athermal dynamics leads to hysteresis and has been extensively studied in the case of the RFIM \cite{sethna93,dahmen96,perkovic,perez04}. For the $O(N \geq 2)$ model a different drive has also been considered in which the driving force is not an applied conjugate source $\bm J_{t}$ but is equal to $v \partial_x \bm\varphi_{xt}$ where $v$ is a finite driving velocity \cite{moving-glass,haga}.

\section{Brief recap of results prior to 2004}
\label{sec_recap}

In this section we briefly summarize the equilibrium properties of  the RFIM and its $O(N)$ extension that were established by 2004 (which is the publication year of our first nonperturbative FRG paper \cite{tarjus04}). Before 2004, there has also been an extensive body of work on the behavior of the RFIM when quasi-statically driven at zero temperature, which was introduced by Sethna and coworkers \cite{sethna93,dahmen96,perkovic,perez04}. The physics then involves hysteresis, avalanches, and out-of-equilibrium criticality, and its study sheds some interesting light on the equilibrium behavior itself. However, we will not dwell on it here.

In the RF$O(N)$M in equilibrium, there is a transition between a paramagnetic phase (at high temperature and large disorder strength) and a ferromagnetic phase (at low temperature and small disorder strength) via a critical point for all dimensions $d$ above some lower critical dimension $d_{lc}$. For the short-ranged models $d_{lc}=4$ for $N>1$ and $d_{lc}=2$ for $N=1$. Both values of $d_{lc}$ have been subject to contention for some time. In the case of the RFIM ($N=1$) a heuristic argument put forward by Imry and Ma \cite{imry-ma75} suggested that an infinitesimal amount of disorder destabilizes the ferromagnetic phase for dimensions smaller than two, pointing to $d_{lc}=2$. In contrast, perturbation theory\cite{aharony76,grinstein76,young77} at all orders as well as an argument invoking an underlying supersymmetry of the model at zero temperature\cite{parisi79} predicted that a property of dimensional reduction, namely that the critical behavior of the RFIM in dimension $d$ is the same as that of the pure Ising model in dimension $d-2$, which implies a lower critical dimension of $3$. Rigorous results have definitely settled the issue in favor of the Imry-Ma prediction with $d_{lc}=2$ \cite{imbrie84,bricmont87,aizenman-wehr}. A review on the theory of the RFIM before 1997 can be found in Ref. [\onlinecite{nattermann98}].

For models with a continuous $O(N)$ symmetry both the Imry-Ma argument and the dimensional-reduction one predict that $d_{lc}=4$ for the paramagnetic to ferromagnetic. Beside the fact that this does not guarantee that these approaches are valid, there remains the possibility of having a transition to a system with quasi-long-range order (QLRO) instead of the conventional long-ranged order \cite{giamarchi_bragg,blatter_review,giamarchi_bragg-review}. Rigorous results have shown that this cannot take place for $N\geq3$ \cite{feldman_exact} but the issue of the lower critical dimension of QLRO for $1<N<3$, which includes the physical value of $N=2$ for which it has been argued that a ``Bragg glass" phase with QLRO is present in $d=3$ \cite{giamarchi_bragg,gingras_bragg}, was still pending.

Above the lower critical dimension, there is strong evidence that the equilibrium critical behavior of the RF$O(N)$M is controlled by a {\it zero-temperature fixed point} \cite{villain84,fisher_activated,bray-moore_RFIM}. This is a new type of fixed point at which the renormalized temperature is irrelevant, albeit ``dangerously" so, and is characterized by a new exponent $\theta>0$. This exponent $\theta$ is equal to $2$ in the mean-field limit. Below the upper critical dimension, which is found equal to $d_{uc}=6$ for the RF$O(N)$M by considering perturbation theory and the Ginzburg criterion, the exponent $\theta$, just like the other critical exponents, may take nontrivial values depending on the dimension $d$. The fact that the critical behavior is controlled by a fixed point at zero (renormalized) temperature and the core of the above mentioned Imry-Ma argument that involves a competition of interactions with no consideration of entropy are the signature that the long-distance behavior of random-field models is dominated by the fluctuations induced by the quenched disorder rather than by thermal fluctuations. As a result, the critical behavior can be directly investigated at zero temperature with the disorder strength as the main control parameter, through the study of the ground state properties.

As a consequence of the ``dangerous irrelevance" of temperature, the scaling behavior at criticality is characterized by a modified hyperscaling relation, $2-\alpha=(d-\theta)\nu$, where as usual $\alpha$ is the specific-heat exponent and $\nu$ the correlation-length exponent. There are also two pair correlation functions and two ``anomalous dimensions" of the field at criticality, with 
\begin{equation}
\begin{aligned}
  \label{eq_propag_conn}
  G_{\rm{conn}}(x-y)&=\overline{\langle \varphi(x)\varphi(y)\rangle}-\overline{\langle\varphi(x)\rangle \langle\varphi(y)\rangle}
  \\&\sim \frac {T}{\vert x-y\vert^{d-2+\eta}}
\end{aligned}
\end{equation}
the so-called ``connected" pair correlation function and
\begin{equation}
\begin{aligned}
  \label{eq_propag_disc}
  G_{\rm{disc}}(x-y)&=\overline{\langle \varphi(x)\rangle \langle\varphi(y)\rangle} - \overline{\langle \varphi(x)\rangle}\; \overline{\langle\varphi(y)\rangle}
  \\& \sim \frac {1}{\vert x-y\vert^{d-4+\bar\eta}}
\end{aligned}
\end{equation}
the so-called ``disconnected" correlation function, where we have considered the case $N=1$ for simplicity. In the above equations, $\langle\cdot\rangle$ denotes the thermal average and  an overline the average over the random field. The connected correlation function measures the influence of thermal fluctuations (and vanishes at zero temperature) whereas the disconnected one is sensitive to the fluctuations of the quenched disorder, i.e., the sample-to-sample fluctuations, and diverges more strongly at the critical point. The two anomalous dimensions are related by an expression involving the temperature exponent as
\begin{equation}
  \label{eq_eta-etabar}
\bar\eta -\eta =2- \theta\,.
\end{equation}
At the upper critical dimension ($d_{uc}=6$), $\bar\eta=\eta=0$ and $\theta=2$, whereas at the lower critical dimension of the RFIM, $d_{lc}=2$, one expects that $\bar\eta=2\eta=2$ and $\theta=1$ \cite{bray-moore_RFIM}. (In the RF$O(N>1)$M at the lower critical dimension for long-range ferromagnetism, $d_{lc}=4$, one finds $\bar\eta=\eta=0$ and $\theta=2$.)

The dangerous irrelevance of the temperature shows up in the slowing down of dynamics when approaching the critical point. In the case of the RFIM, the latter takes an activated dynamical scaling form in which it is the logarithm of the relaxation time $\tau$ that grows as a power law of the correlation length $\xi$,
\begin{equation}
  \label{eq_activated_scaling}
\log \tau \sim \xi^\psi
\end{equation}
with $\psi$ some {\it a priori} unknown exponent \cite{villain84,fisher_activated}, instead of the form $\tau \sim \xi^z$ found in conventional critical slowing down (formally, $z=\infty$ here) .

Ever since the introduction of the model the equilibrium behavior of the RFIM on Euclidean lattices has been extensively studied by computer simulation, mostly in $d=3$. Large-scale computer simulations can be performed at $T=0$ where combinatorial algorithms allow one to find the (almost surely) unique ground state of a finite sample in polynomial time \cite{rieger_algo}. By using system sizes up to $256^3$ spins and a careful finite-size scaling analysis, Middleton and Fisher then unambiguously showed that the transition in $d=3$ is a critical, second-order one for a Gaussian distribution of the random fields \cite{middleton-fisher02}.

Note finally that there have also been attempts to explain the breaking of dimensional reduction and of the underlying supersymmetry of the zero-temperature construction below the upper critical dimension, mostly within the replica formalism. Instantons in replica space \cite{dotsenko}, bound states between replicas associated with the putative divergence of some Bethe-Salpeter kernel \cite{dedom-orland_BS,brezin-dedom_BS,sourlas_BS}, some replica symmetry breaking mechanism \cite{mezard-young}, etc., have been invoked based on partial, usually perturbative, calculations but have not been conclusive.

To conclude this section one can list a number of unresolved questions at the time: What is the mechanism for the breakdown of dimensional reduction and the breaking of the underlying SUSY below $d=6$? Can one provide a theoretical computation of the critical exponents, in particular of the exponent $\psi$ characterizing the activated dynamic scaling? Is it possible to theoretically describe collective phenomena such as avalanches and droplets? Is the critical scaling described by 2 or 3 independent exponents? What is the phase behavior of the RF$O(N)$M in the whole ($N$, $d$) plane and what is the lower critical dimension of QLRO for $N=2$? Are the equilibrium and out-of-equilibrium critical points of the RFIM in the same universality class? These are questions that will be answered by the FRG approach.

\section{Zero-temperature fixed points, avalanches and droplets: The need for a functional RG}
\label{sec_avalanches}

\subsection{Metastable states, avalanches and droplets in the RFIM}

The presence of quenched disorder generically leads at zero temperature to a multiplicity of metastable states, i.e., minima of the bare action that satisfy the stochastic field equation in Eq. (\ref{eq_stochastic_GS}) or, equivalently, of minima of the lattice Hamiltonian in Eq. (\ref{eq_latticeRFO(N)M}). This multiplicity is known for instance to invalidate the straightforward implementation of the supersymmetric formalism that assumes a unique solution of the stochastic field equation \cite{parisi84}. In the case of the RFIM metastable states are generically found in a whole region of the the magnetization/applied-field diagram(to use again the language of magnetic systems) \cite{guagnelli_MS,mlr_MS}.

Associated with the presence of metastable states is another important phenomenon. In any finite sample of, say a RFIM, the ground state is almost certainly unique when the distribution of the random field is continuous. However, when considering the evolution of the ground state under a change of the applied source, one observes abrupt switches at a set of discrete, sample-dependent, values of the source. (Exactly at these specific values there is a coexistence between two ground states, but for infinitesimal changes in one direction or another one state becomes of lower energy and the other one is then ``metastable".) These events have been observed in computer simulation at zero temperature \cite{vives_GS,wu-machta_GS,liu_GS} and are called ``static'' avalanches by analogy with the ``dynamic" avalanches that take place out of equilibrium, between two metastable states of the system, when the RFIM is quasi-statically driven by the external source \cite{sethna01,sethna93,dahmen96,perkovic,perez04}. The same phenomenon of avalanches is also seen in the behavior of an elastic manifold in a random environment, both in equilibrium when the system is in the pinned phase \cite{static-aval} (static avalanches, also referred to as shocks \cite{BBM}) and out of equilibrium at the depinning transition (dynamic avalanches) \cite{depin-aval}.

\begin{figure}[tb]
 \begin{center}
 \includegraphics[width=.5\linewidth]{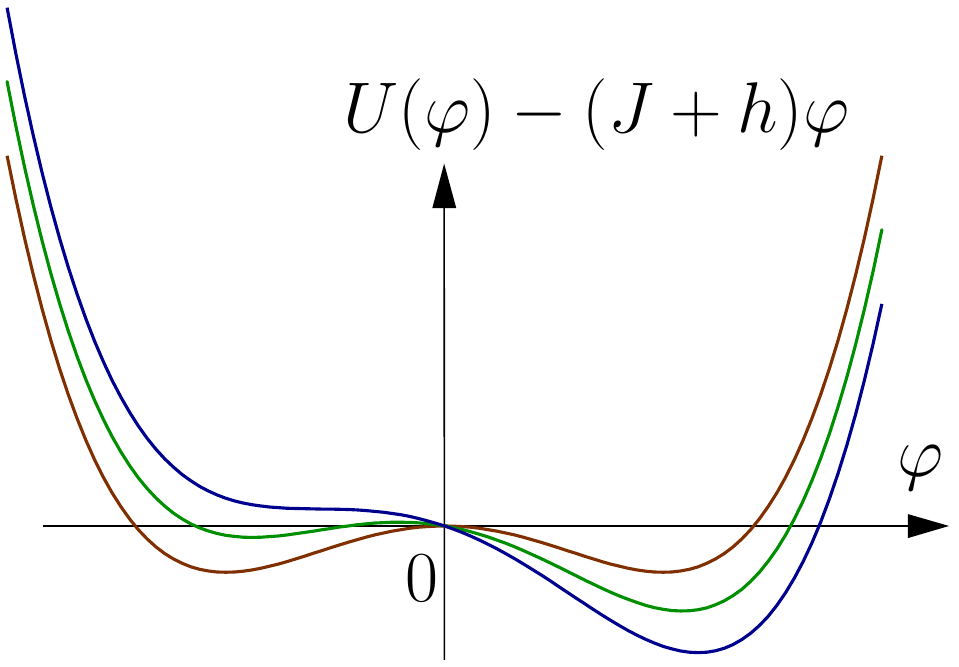} 
  \includegraphics[width=.5\linewidth]{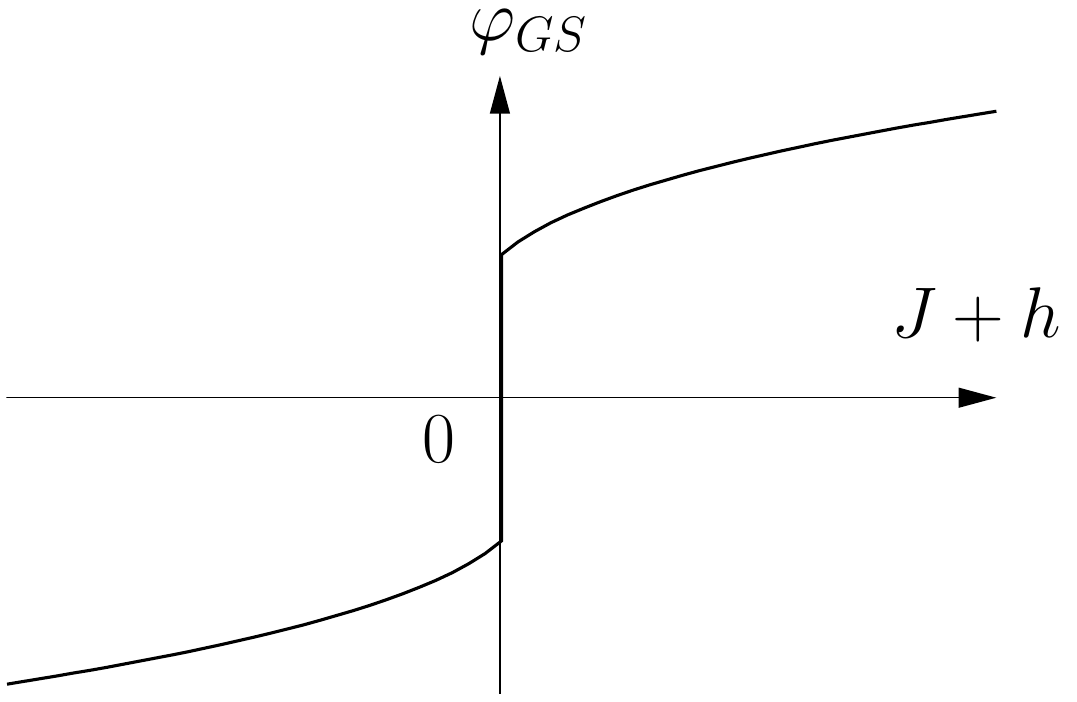} 
   \includegraphics[width=.5\linewidth]{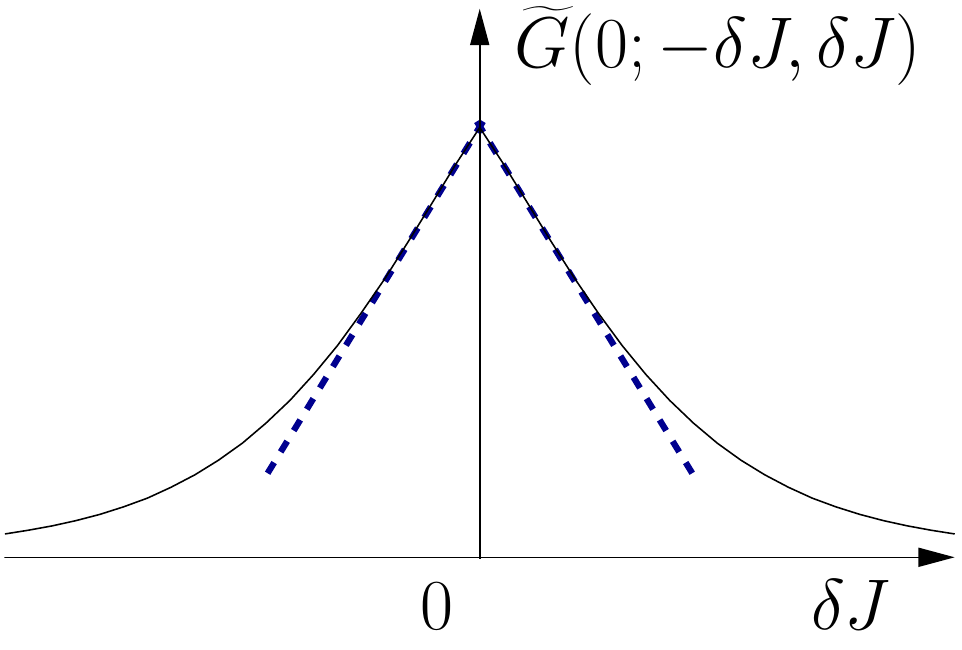}  
    \caption{Illustration  of  avalanches  and of their  consequence on the functional dependence of a disorder-averaged correlation function in the toy model  of the $d=0$ RFIM studied in equilibrium at $T= 0$ (see main text).   (a) Potential $U(\varphi)-(J+h)\varphi$ versus $\varphi$ for different values of $J$, with $U(\phi) =-(\vert r\vert/2)\varphi^2+ (u/4!)\varphi^4$; (b) Ground state configuration $\varphi_{GS}(J+h)$ associated with (a). (c) Two-point correlation function $\widetilde G(0;-\delta J, \delta J)) =\overline{\phi_{GS}(-\delta J+h)\phi_{GS}(\delta J+h)}-\overline{\phi_{GS}(-\delta J+h)}\;\overline{\phi_{GS}(\delta J+h)}$, where the average is over a Gaussian distributed random field $h$. Notice the linear cusp around $\delta J=0$.}
    \label{fig_0d-RFIM}
  \end{center}
\end{figure}

The fact that abrupt changes corresponding to discontinuous variations of the magnetization are found at zero temperature should come as no surprise. In disordered systems, this can take place even in noninteracting zero-dimensional models. Consider for instance a $d=0$ (single point) $\varphi^4$ theory with parameters such that the potential has two minima and the field $\varphi$ is coupled to a random source $h$, which is Gaussian distributed with variance $\Delta_B$, and to a controllable source $J$, i.e., $S(\varphi;h+J)= - (\vert r\vert/2) \varphi^2 + (u/4) \varphi^4-(h+J)\varphi$. Then, according to the value of $h+J$, the ground state of the system will switch from the vicinity of one minimum to that of the other one with a jump when $h+J=0$. This jump, whose location is sample ($h$) dependent, corresponds to an avalanche, albeit a zero-dimensional one. This is sketched in Fig. \ref{fig_0d-RFIM} (a) and (b).

Droplets on the other hand are rare low-energy excitations having an energy difference with the ground state that can be as small as wanted. In particular, this difference can be smaller than the temperature, whatever the nonzero value of the latter. The existence of such droplets has been postulated in phenomenological approaches \cite{fisher_activated,SG_droplet} and has found support in simulations of the RFIM \cite{hartmann_droplet}. Although rather trivial, the $d=0$ RFIM introduced above illustrates what a ``droplet" can be: When the two minima of the action are almost degenerate, their contribution to the partition function even at a very low (but nonzero) temperature becomes comparable, since the Boltzmann weight of the ground state no longer dominates that of the ``metastable" state when the difference in energy is or the order or less than the temperature $T$. In finite, nonzero dimension $d$ such a situation rarely occurs for states that differ on large length scales, but it has been conjectured that thermally active (i.e., quasi-degenerate with the ground state) droplets appear on a large size $L$ with a power-law decaying probability $\propto T L^{-\theta}$, with $\theta$ the temperature exponent \cite{fisher_activated,SG_droplet}.  

As will be illustrated in more detail below, these avalanches and droplets generate singular functional dependences in the disorder-averaged correlation functions and 1PI disorder cumulants. However, for avalanches and droplets to affect the long-distance physics of a $d$-dimensional disordered model with $d>0$, they must be of collective origin and occur on all scales (unlike in the $0$-dimensional toy model discussed above).

Note finally that avalanches (and droplets) are in general harder to characterize in the case of the continuous $O(N)$ symmetry because of the many directions in which they can extend, but they are nonetheless present.

\subsection{The need for multiple copies}

Whether one study random-field systems in or out of equilibrium, the central quantities are generating functionals, as, e.g., the equilibrium ``free energy'' functional $\mathcal W[\bm J;\bm h]$ previously introduced (Sec. \ref{sec_models}). In the presence of quenched disorder, such functionals are random, i.e., sample dependent. Therefore, they are fully characterized by their (functional) probability distribution or, alternatively, by the infinite set of their cumulants (if of course the cumulants exist). Dealing with cumulants has the advantage of involving an average over the bare disorder: As a result, one recovers the translational and rotational invariances in Euclidean space which are otherwise broken by the space-dependent random field. We will thus consider a formalism based on cumulants. However, a crucial point when working with such disorder-averaged quantities is that one does not want to lose track of the rare or singular collective phenomena (avalanches and droplets) taking place in the system's samples and discussed just above.

To illustrate the effect of avalanches and droplets on disorder-averaged quantities, we consider again the case of the $d=0$ RFIM. Let study first the case of  zero temperature, $T=0$. Consider two copies of the system with the same disorder $h$ but submitted to different sources $J_1=J+\delta J$ and $J_2=J-\delta J$ and compute the correlation function $\widetilde G(J_1,J_2)=\overline{\varphi_{GS}(J_1;h) \varphi_{GS}(J_2;h)}-\overline{\varphi_{GS}(J_1;h)}\;\overline{\varphi_{GS}(J_2;h)}$. This is an extension to general sources $J_1\neq J_2$ of what is called the $2$-point ``disconnected" correlation function [see, e.g., Eq. (\ref{eq_propag_disc})]. A simple calculation shows that  when $\delta J\to 0$ this correlation function, which is symmetric under the inversion $\delta J\to -\delta J$, behaves as
\begin{equation}
  \label{eq_0dcusp}
\widetilde G(J+\delta J,J-\delta J)=\widetilde G(J,J) - \frac{24 \;{\rm e}^{-\frac{J^2}{2\Delta_B}}}{u\sqrt{2\pi}\Delta_B}\vert \delta J\vert +{\rm O}(\delta J^2),
\end{equation}
i.e., displays a linear cusp in $\delta J=(J_1-J_2)/2$: see Fig. \ref{fig_0d-RFIM} (c). This nonanalytic dependence on the replica sources is a direct consequence of the avalanches in the ground state. Through a Legendre transform it translates into a cusp in the dependence on the average replica fields of the associated $1$-particle irreducible (1PI) correlation function, which in this case is the second cumulant of the renormalized random field.

If temperature is nonzero, $T>0$, but small, the equilibrium properties now essentially involves a Boltzmann average over the two minima, which form a two-level system. The nonanalyticity is then rounded,
\begin{equation}
  \label{eq_0dBL}
\widetilde G(J+\delta J,J-\delta J)-\widetilde G(J,J) = T f(J,\frac{\delta J^2}{T^2}) + {\rm O}(T^2,\delta J^2)\,,
\end{equation}
where $f(J,y)=f_2(J)y + {\rm O}(y^2)$ when $y\to 0$ and $f(J,y)\sim f_\infty(J) \sqrt y$ when $y\to \infty$. As $f_\infty(J)= 24 \;{\rm e}^{-J^2/(2\Delta_B)}/(u\sqrt{2\pi}\Delta_B)$, one recovers Eq. (\ref{eq_0dcusp}) when $T\to 0$. For  $T>0$ the cusp is rounded in a region where $\vert \delta J\vert\lesssim T$, which shrinks as $T\to 0$. The limit $T\to 0$ is therefore nonuniform in $\delta J$ and involves a ``thermal boundary layer" (see Refs. [\onlinecite{chauve_creep,balents-doussal_BL,ledoussal10}] for the same phenomenon in the case of an interface in a disordered environment).

Generically, in any dimension, avalanches at zero temperature generate cusps in the functional dependence on the field arguments of the cumulants of the renormalized random source and droplets at low but nonzero temperature generate a thermal rounding of these cusps in a boundary layer. Describing such features therefore requires the functional dependence of the cumulants for \textit{generic arguments}. For instance, a complete characterization of the random functional $\mathcal W[ \bm J;\bm h]$ implies the knowledge of all its cumulants, $W_1[ \bm J_1]$, $W_2[ \bm J_1,  \bm J_2]$, $W_3[ \bm J_1,  \bm J_2,  \bm J_3]$, ..., which are defined as
\begin{equation}
  \label{eq_cumW1}
W_1[ \bm J_1]= \overline{\mathcal W[ \bm J_1;\bm h]}
\end{equation}
\begin{equation}
\begin{aligned}
  \label{eq_cumW2}
W_2[ \bm J_1,  \bm J_2]= &\overline{\mathcal W[ \bm J_1;\bm h]\mathcal W[ \bm J_2;\bm h]}\\&-\overline{\mathcal W[ \bm J_1;\bm h]}\,\, \overline{\mathcal W[ \bm J_2;\bm h]},
\end{aligned}
\end{equation}
etc.  

\textit{Generic}, \textit{i.e.} independently tunable, arguments require the introduction of copies or replicas of the original system, each with the same bare disorder (random field) but coupled to \textit{distinct and independent} external sources $\bm J_1$, $\bm J_2$, etc. It is worth stressing that this is \textit{not} what is done in the conventional replica trick \cite{book_SG} nor in the Parisi-Sourlas supersymmetric approach \cite{parisi79}. In the simple implementation of the former, the sources acting on the replicas are all taken equal and in the latter a single copy of the system is considered. As a result, in both cases, one only has access to cumulants in which all the arguments are equal. Quite differently in the present formalism, we consider multiple copies or replicas and sources that explicitly break the (permutational) symmetry among these replicas.

\subsection{Multi-copy formalism}
\label{sec:multi-copy}

The cumulants of the random free-energy functional $\mathcal W[ \bm J;\bm h]$ can be generated from an average involving copies (or replicas) of the original disordered system, as follows:
\begin{equation}
\begin{split} 
 \label{eq_cumW}
&\overline{{\rm exp}( \sum_{a}\mathcal W[\vect J_a; \vect h])} = {\rm exp} \left( W[\left\lbrace \vect J_a\right\rbrace ])\right)  \\& = {\rm exp}\bigg (\sum_{a} W_1[\vect J_a] +\dfrac{1}{2}\sum_{a,b} W_2[\vect J_a, \vect J_b]\\&+ \dfrac{1}{3!}\sum_{a,b,c}W_3[\vect J_a, \vect J_b, \vect J_c] + \cdots \bigg),
\end{split}
\end{equation}
where, as stressed above, the copies have the \textit{same} disorder but are coupled to {\it distinct} external sources. A convenient trick to extract the cumulants with their full functional dependence is to let the number of replicas be arbitrary and to then view the expansion  of the functional $W[\left\lbrace \vect J_a\right\rbrace ]$ in the right-hand side of Eq. (\ref{eq_cumW}) as an expansion in increasing number of unconstrained, or  ``free'', sums over replicas. The term of order $p$ in the expansion is a sum over $p$ replica indices of a functional depending exactly on $p$ replica sources, this functional being precisely equal here to the $p$th cumulant of $\mathcal W[\vect J;\vect h]$. This procedure, in which the permutational symmetry between replicas is explicitly broken, leads to well-defined algebraic manipulations \cite{tarjus04,tarjus08,tissier12,doussal-wiese_replicas,mouhanna_replicas}.

The central object of our FRG approach is not the free-energy functional $W[\{\bm J_a\}]$ but rather its Legendre transform, the effective action $\Gamma[\{\bm \phi_a\}]$, defined by
\begin{equation}
\label{eq_legendre_gamma}
\Gamma[\{\vect \phi_a\} ] = - W[\{\vect J_a\}] + \sum_a \int_{x} \vect J_a(x)\cdot \vect \phi_a(x),
\end{equation}
where
\begin{equation}
\label{legendre_phi}
\phi_a^\mu(x)=\frac{\delta W[\{\vect J_e\}] }{\delta  J_a^\mu(x)}
\end{equation}
is the classical or average field, $\bm \phi_a(x)=\langle\bm \varphi_a(x)\rangle$. $\Gamma[\{\bm \phi_a\}]$ is the generating functional of the 1PI correlation functions and in the language of magnetic systems it represents a Gibbs free-energy functional while the $\bm \phi_a(x)$'s are the local magnetizations.

The effective action  $\Gamma [\left\lbrace \vect \phi_a\right\rbrace ]$ can also be expanded in increasing number of free replica sums,
\begin{equation}
\begin{split} 
 \label{eq_cumg}
\Gamma [\left\lbrace \vect \phi_a\right\rbrace ]= & \sum_{a}\Gamma_1[\vect \phi_a] -\dfrac{1}{2}\sum_{a,b}\Gamma_2[\vect \phi_a, \vect \phi_b]\\& + \dfrac{1}{3!}\sum_{a,b,c} \Gamma_3[\vect \phi_a, \vect \phi_b, \vect \phi_c] + \cdots,
\end{split}
\end{equation}
where we have purposedly introduced a minus sign for all even terms of the expansion. The $\Gamma_p$'s and the $W_p$'s are related through the Legendre transform and a term-by-term identification of the expansions in free replica sums. $\Gamma_{p=1}$ is the disorder-averaged effective action and, with a grain of salt \cite{tarjus08,tissier12}, the $\Gamma_p$'s for $p \geq 2$ can be considered as ``cumulants of the renormalized or effective disorder". Their functional derivative $\Gamma_{p;x_1\mu_1,x_2\mu_2\cdots,x_p\mu_p}^{(11\cdots1)}[\vect \phi_1, \vect \phi_2, \cdots, \vect \phi_p]$ can then be viewed as ``cumulants of the renormalized or effective random field". (Here and below, superscripts with parentheses denote the order of the functional derivatives with respect to the appropriate arguments.) The knowledge of the complete set of these cumulants, with generic arguments, fully characterizes the theory.

\section{Summary of FRG results}
\label{sec_summary}

\subsection{Equilibrium criticality: The way out of dimensional reduction and the spontaneous breaking of SUSY}

\begin{figure}[tb]
   \begin{center}
\includegraphics[width=\linewidth]{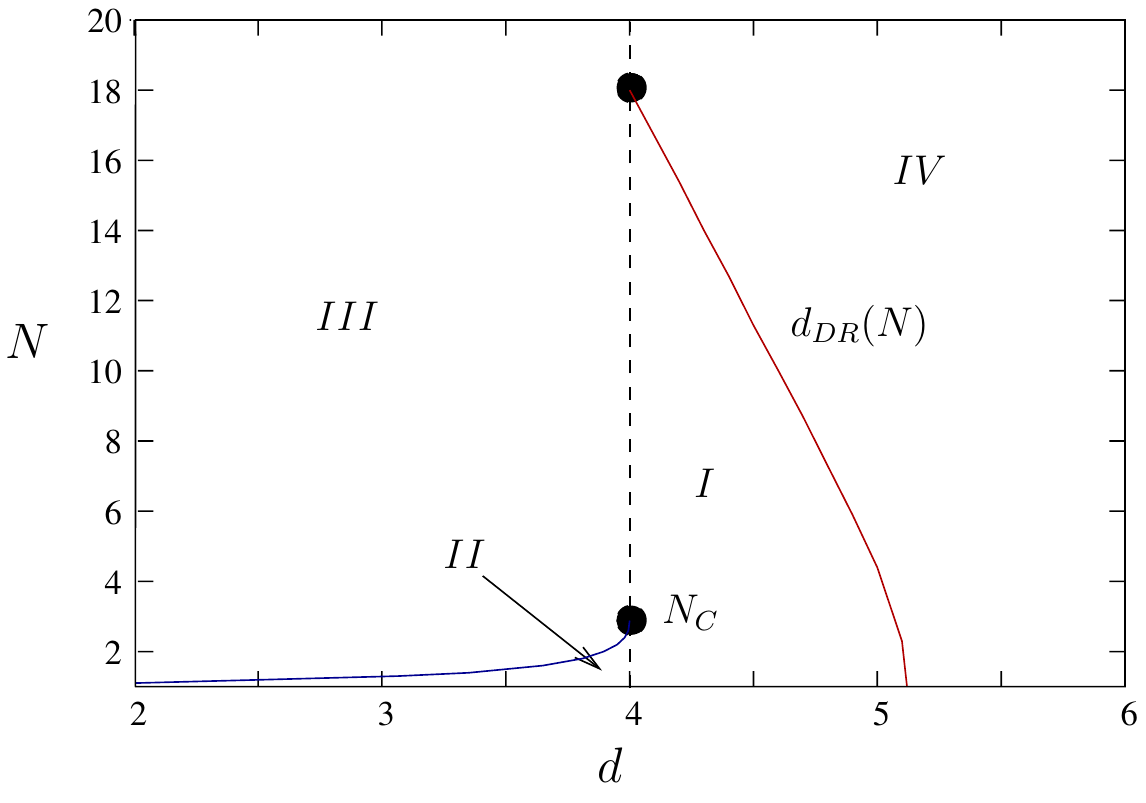}
\caption{Nonperturbative FRG prediction of the equilibrium phase behavior of the $d$-dimensional RF$O(N)$M. In region $III$, there are no phase transitions and the system is always disordered (paramagnetic). In regions $I$ and $IV$ , there is a second-order paramagnetic-to-ferromagnetic transition and in region $II$, a second-order transition between paramagnetic and QLRO phases. In region $IV$ the nonanalyticity of the dimensionless effective action at the zero-temperature fixed point is weak enough that the critical exponents take their dimensional-reduction value and SUSY is valid, whereas a complete breakdown of dimensional reduction and a concomitant breaking of SUSY take place in regions $I$ and $II$. Regions $I$ and $IV$ are separated by a nontrivial critical line $d_{DR}(N)$. Above $d=6$, the critical behavior is described by classical (mean-field) exponents. Note that the baseline corresponds to $N=1$, i.e., to the RFIM.}
\label{fig_RFO(N)_phase_diagram}
\end{center}
\end{figure}

The main outcome of our FRG  investigations concerning the equilibrium critical behavior of the RF$O(N)$M is the existence of a critical line $d_{DR}(N)$ separating  in the ($d,N$) plane a domain above the line in which the the main scaling behavior at the critical point is given by the $d\to d-2$ dimensional-reduction property and below which this dimensional reduction breaks down \cite{tarjus04,tissier12,tissier06,tissier11}. The critical line, which is plotted in Fig. \ref{fig_RFO(N)_phase_diagram}, starts near $d_{DR}(N=1)\approx 5.1$ for the Ising version and reaches $d_{DR}=4$ for $N\approx 18$. This result explains how one goes from the upper critical dimension $d_{uc}=6$ in the vicinity of which dimensional reduction is valid to low dimensions such as $d=3$ where, in accord with rigorous results,  it is broken. It is obtained via a nonperturbative implementation of the FRG that allows us to compute the nontrivial critical dimension $d_{DR}(N)$ as well as critical exponents and fixed-point functions. It is furthermore supported by perturbative FRG approaches for the $O(N>1)$ version in $d=4+\epsilon$ at one- and two-loop levels \cite{tarjus04,tissier06,tissier_2loop} and for the RFIM in $d=6-\epsilon$ at two loops when considering nonanalytic functional ``cuspy'' perturbations around the cuspless Gaussian fixed point on top of the usual irrelevant directions \cite{tarjus_perturb}.

In the FRG context the breakdown of dimensional reduction is associated with the appearance of a strong nonanalytic dependence (a cusp) on the arguments \cite{footnote_arguments} of the cumulants of the renormalized random field at the zero-temperature fixed point, similarly to what previously found in the perturbative FRG of an elastic manifold in a random environment \cite{fisher86b,nattermann,narayan92,FRGledoussal-chauve,doussal_2loop,FRGledoussal-wiese}.  The critical dimension $d_{DR}(N)$ corresponds to the point where the ``cuspless" fixed point give way to the ``cuspy" fixed point. Actually, this occurs through different mechanisms for large and for small $N$. For the RFIM the cuspless fixed point associated with dimensional reduction disappears at $d_{DR}\approx 5.1$ and the cuspy fixed point emerges continuously below $d_{DR}$ through a boundary-layer mechanism \cite{baczyk_FP}. This explains the unusual properties of the corrections to scaling in the RFIM below $d_{DR}$ \cite{balog_comment19}. (Note also that this theoretical explanation of dimension-reduction breakdown is fully compatible with the rigorous proof that no {\it bona fide} spin-glass phase \cite{krzakala_RFIMSG} nor spontaneous replica-symmetry breaking \cite{noRSB_math} can exist in the RFIM.)

Within the framework of a superfield and superspace formulation, the nonperturbative FRG also provides an explanation for the breaking of the underlying SUSY \cite{tissier11,tissier12,tissier_unpublished}. Above $d_{DR}(N)$ SUSY, which is a rotation invariance in superspace, is valid at the fixed point and, even if one starts with a non-SUSY initial condition, it is restored at large distance along the FRG flow. On the other hand, SUSY is broken at the fixed point below $d_{DR}(N)$. If one initiates the FRG flow with a SUSY condition, one finds a spontaneous SUSY breaking at a finite scale along the flow. This SUSY breaking is associated with the appearance of cusps in the functional field dependence of the renormalized cumulants, cusps that lead to a breakdown of the SUSY Ward identities \cite{tissier11,tissier12}. The scenario of a restoration of SUSY and dimensional reduction above some dimension close to $5$ for the RFIM is supported by recent large-scale computer simulations \cite{fytas_5d,fytas_SUSY}.

\subsection{Physical interpretation: avalanches at zero temperature and droplets at low temperature}

\begin{figure}[ht]
\centering
\includegraphics[width=\linewidth]{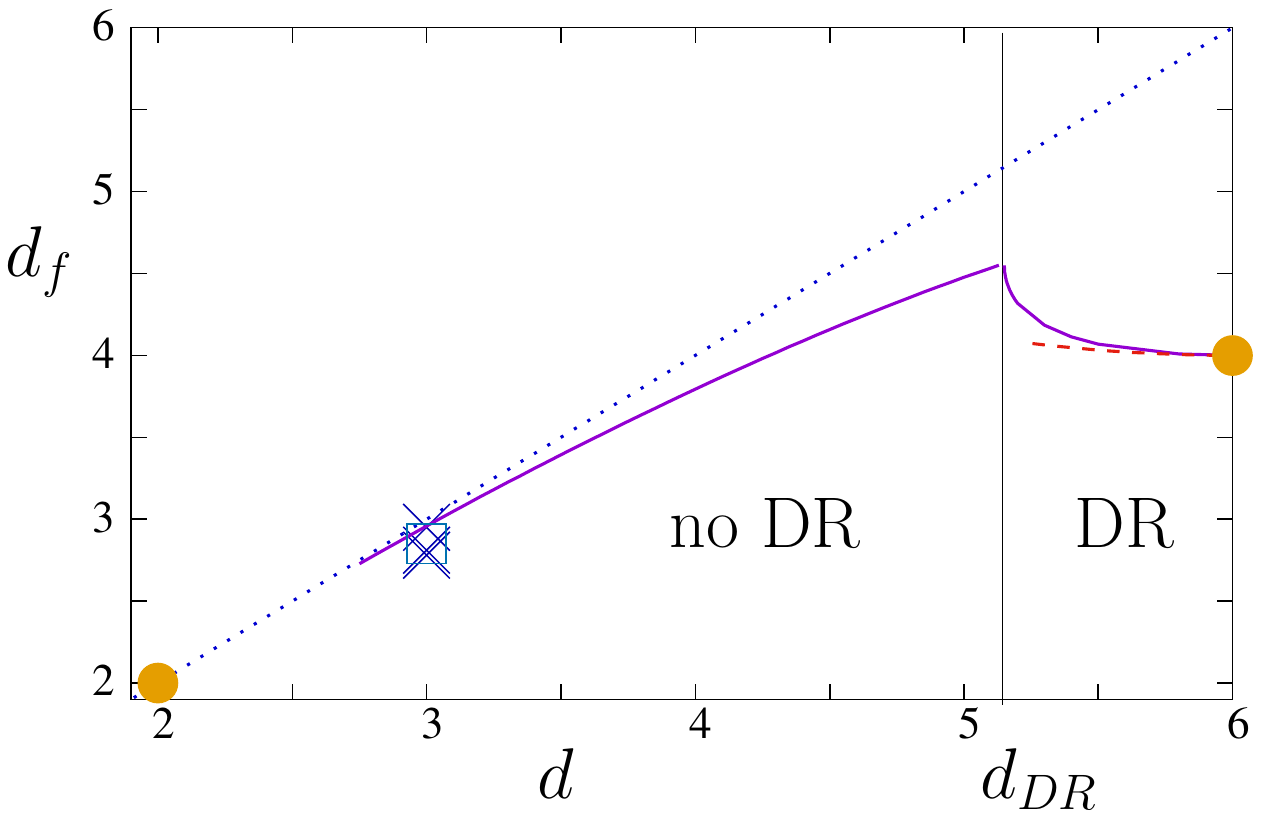}
\caption{Nonperturbative FRG prediction of the fractal dimension $d_f$ of the largest typical avalanches versus $d$ for the RFIM at the equilibrium critical point. The filled circles indicate the known values at $d_{lc}=2$ and $d_{uc}=4$. The dotted line is the upper bound $d_f\leq d$. Below $d_{DR}\approx 5.1$, $d_f$ is equal to $(d+4-\bar\eta)/2$ . The red dashed curve corresponds to the two-loop perturbative FRG calculation calculation in $\epsilon =6-d$, $d_f= 4+(7/54) \epsilon^2+{\rm O}(\epsilon^3)$ \cite{tarjus_perturb}. The symbols are estimates obtained from computer simulations of the RFIM in $d=3$ either in equilibrium \cite{liu09} or out of equilibrium \cite{perkovic,perez04,liu09}. (Note that the numerical resolution of the FRG flow equations becomes extremely difficult for the RFIM in low dimension, typically for $d\lesssim 2.9$, so that we have no results there.)}
\label{fig_df}
\end{figure}

As stressed in Sec. \ref{sec_avalanches}, the nonanalytic field dependences of the cumulants of the renormalized random field are generated by the presence of abrupt collective phenomena, described as avalanches (or shocks), in the evolution of the ground state as a function of the applied source. At zero temperature, avalanches on all scales are always present. This is seen for instance in the mean-field limit where the avalanche properties can be exactly computed. However, when avalanches and the resulting cusps are subdominant in an RG sense near the zero-temperature fixed point, dimensional reduction and SUSY are still satisfied at this fixed point. The fractal dimension $d_f$ of the largest typical critical avalanches is then smaller than the fractal dimension of the total order parameter, $(d+4-\bar\eta)/2$, and there is a diverging number of such avalanches at criticality, which is characterized by the exponent \cite{tarjus13}
\begin{equation}
\label{eq_lambda_df}
\lambda=\frac{d+4-\bar\eta}{2}-d_f  >0\,,\;\; {\rm for}\; d\geq d_{DR}\,.
\end{equation}
These exponents, $\lambda$ and $d_f$, can be computed through the nonperturbative FRG \cite{tarjus13,baczyk_FP} and the perturbative FRG in $\epsilon=6-d$ \cite{tarjus_perturb}. On the other hand below $d_{DR}$, avalanches and cusps dominate the fixed point and the whole critical scaling, so that $d_f=(d+4-\bar\eta)/2$ and $\lambda=0$. The fractal dimension of the largest typical avalanches at criticality is plotted as a function of dimension in Fig. \ref{fig_df}. Note that the same criterion concerning the fractal dimension of the avalanches can be used to rationalize why dimensional reduction is always broken for elastic manifolds in a random environment below their upper critical dimension $d_{uc}=4$ and why it is always valid  for the statistics of dilute branched polymers below the upper critical dimension $d_{uc}=8$ \cite{tarjus13}.

Criticality at a small but nonzero temperature involves the physics of power-law rare excitations known as droplets \cite{fisher_activated}. Within the nonperturbative FRG this is captured through the thermal rounding of the cusps that are present in the renormalized cumulants at the zero-temperature fixed point (whether subdominant or dominant). The renormalized temperature flows to zero but the limit is highly nonuniform in the field-dependent cumulants and proceeds via a ``thermal boundary layer", as first found in the case of the random elastic manifold model \cite{chauve_creep,balents-doussal_BL,ledoussal10}. This manifestation of the dangerous irrelevance of the temperature leads to anomalous thermal fluctuations and activated dynamic scaling in the RFIM that can both be described by the nonperturbative FRG \cite{tissier06,balog_activated}.

\subsection{Unified description of ferromagnetism, quasi-long-range order (QLRO) and criticality in the whole ($N$, $d$) plane}

\begin{figure}[tb]
   \begin{center}
\includegraphics[width=\linewidth]{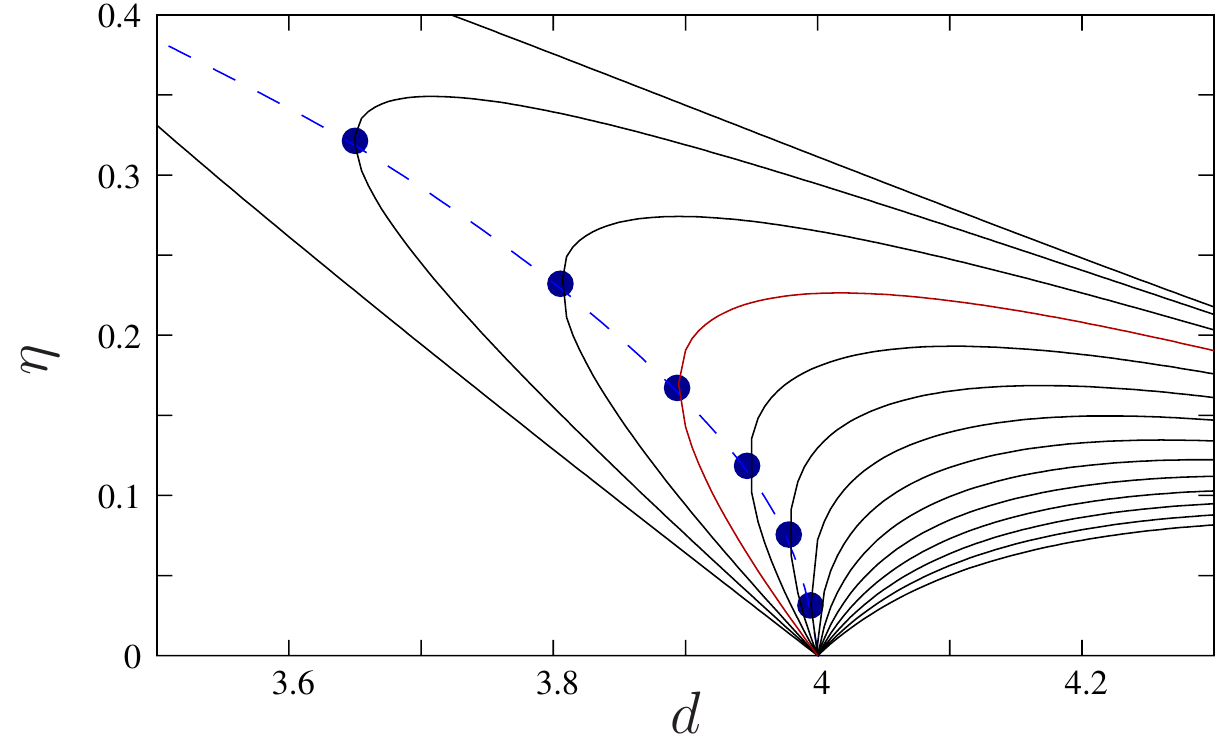}
\caption{QLRO lower critical dimension for the equilibrium RF$O(N)M$ from the nonperturbative FRG:  The anomalous dimension $\eta$ associated with the zero-temperature ``cuspy" fixed points are plotted versus $d$ for values of $N$ ranging from $1.4$ to $4$ by steps of $0.2$. For $N > N_c=2.8347\cdots$, only one fixed-point value emerges from the point ($\eta= 0$, $d = 4$); but for $N < N_c$, one finds two values of $\eta$ for each dimension, the upper one being associated with the critical fixed point and the lower one with the QLRO fixed point. The two branches of fixed points coalesce for a value $d_{lc}(N)$ shown by (blue) filled circles and the dashed line. This value is found around $3.9$ for $N=2$ (red curve).}
\label{fig_eta_QLRO}
\end{center}
\end{figure}

The  nonperturbative FRG approach of the RF$O(N)$M provides a unified picture of ferromagnetism, QLRO and criticality in the whole ($N$, $d$) plane thanks to the property that the resulting flow equations can be solved {\it for any value of the number of components $N$ and the dimension $d$} \cite{tarjus04,tissier06}. We have found that below a critical value $N_c = 2.8347\cdots$ and for $d < 4$ the model has a transition to a QLRO phase. Both this phase and the transition to it (from the paramagnetic phase) are governed by zero-temperature nonanalytic (cuspy) fixed points. The transition disappears below a lower critical dimension $d_{lc}^{{\rm QLRO}}$ which we find around $3.9$ for $N = 2$: see Fig. \ref{fig_eta_QLRO}. Therefore, contrary to previous claims \cite{giamarchi_bragg,gingras_bragg}, no QLRO and no topologically ordered Bragg glass phase exist in the $3$-$d$ RF$XY$M. (One should however be cautious about concluding that no Bragg glass phase can be found in $3$-dimensional physical systems because the description through the simple RF$XY$M may be insufficient.)  The predictions from the  nonperturbative FRG concerning the scenario of dimensional-reduction breakdown/restoration as well as the disappearance of QLRO due to collapse with another zero-temperature fixed point are supported by the analysis through a perturbative FRG to two loops in $d = 4 \pm \epsilon$ \cite{tissier_2loop,doussal-wiese_RF}.

As seen from Fig. \ref{fig_RFO(N)_phase_diagram} the topology of the ($N$, $d$) diagram describing the phase behavior of the RF$O(N)$M is similar to that of the pure $O(N)$ model in $2$ dimensions less, even though the dimensional-reduction property breaks down below the critical line $d_{DR}(N)$ \cite{tissier06,baczyk_FP}. However, through the two-loop perturbative FRG near $d=4$ one finds that the special point ($N_c = 2.8347\cdots$, $d = 4$), which is the analog of the point ($N = 2$, $d = 2$) for the pure $O(N)$ model, does not correspond to a Berezinskii-Kosterlitz-Thouless transition but rather to a conventional second-order transition (the beta function which vanishes at one loop is indeed not identically zero at two loops) \cite{tissier_2loop}.

\subsection{3 independent exponents describe the critical scaling}

Whereas phenomenological theories take the temperature exponent $\theta$ as an independent input,\cite{villain84,fisher_activated} which implies that equilibrium scaling behavior is described by three independent exponents in place of the usual two-exponent scaling for finite-temperature fixed points, Schwartz and coworkers \cite{schwartz_2exp} have claimed that $\theta=2-\eta$, or equivalently $\bar\eta=2\eta$, so that scaling is described by only two independent exponents. Whereas this is indeed verified for the RFIM near the lower critical dimension at first order in $\epsilon=d-2$ \cite{bray-moore_RFIM}, the derivation leading to the conclusion is supposed to hold for the Ising as well as the continuous version with $O(N)$ symmetry, and for all dimensions $d$.

\begin{figure}[tb]
   \begin{center}
\includegraphics[width=\linewidth]{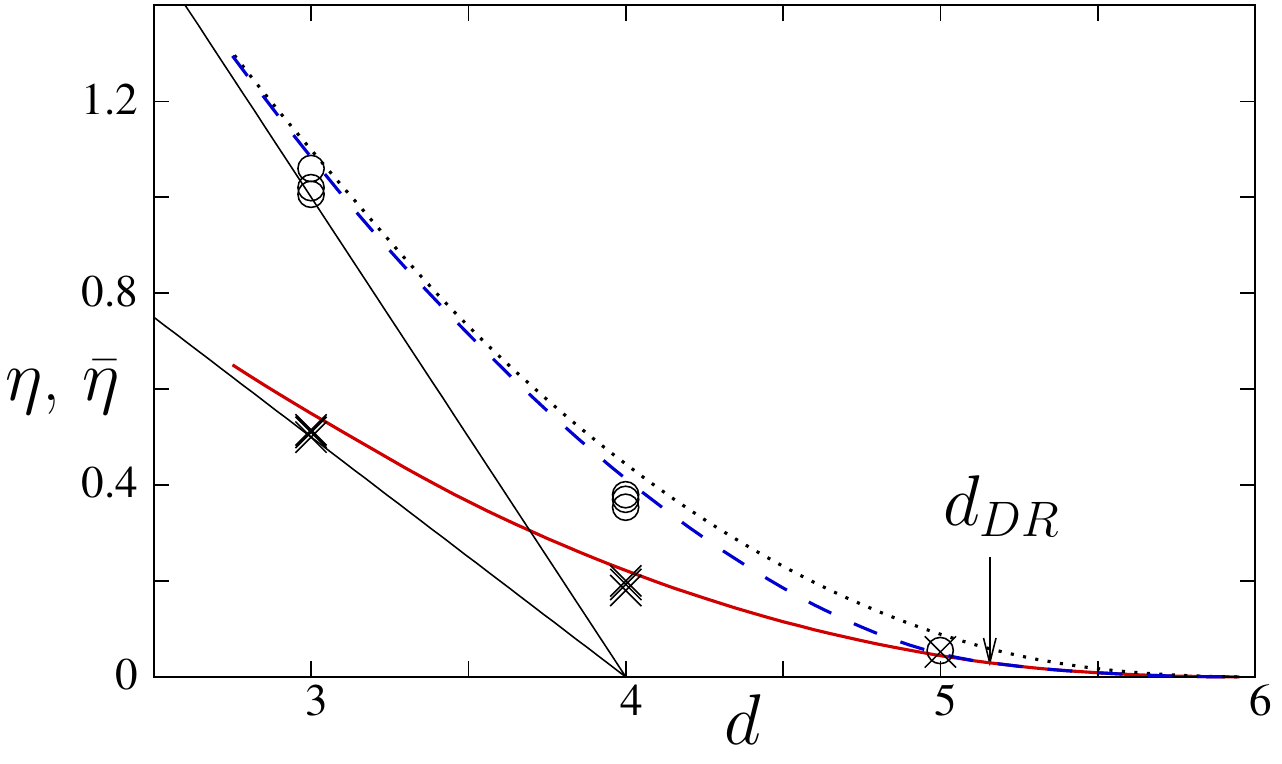}
\caption{Nonperturbative FRG prediction for dependence on the spatial dimension $d$ of the anomalous dimensions $\eta$ (lower, full red curve) and $\bar\eta$ (upper, dashed blue curve) for the equilibrium RFIM at criticality. The full lines are lower bounds for the anomalous dimensions [$(4-d)/2$ for $\eta$ and $4-d$ for $\bar\eta$] and the dotted line is the prediction $\bar\eta=2\eta$ \cite{schwartz_2exp}. The symbols represent the results of large-scale ground-state computations in $d=3$\cite{middleton-fisher02,hartmann_3d,fytas_3d}, $d=4$ \cite{middleton_4d,hartmann_4d,fytas_4d} and $d=5$ \cite{fytas_5d}. The critical dimension above which dimensional reduction is obeyed (with $\eta=\bar\eta$) is $d_{DR}\approx 5.1$.}
\label{fig_etabareta_RFIM}
\end{center}
\end{figure}

Through the nonperturbative FRG and the perturbative FRG near $d=4$ for the RF$O(N>1)$M \cite{tarjus04,tissier06,tissier12,tarjus_exponents}, we have unambiguously shown that the two-exponent scenario cannot be right in general. Indeed, there is a whole region of the ($d$,$N$) plane where dimensional reduction is restored with $\bar\eta=\eta \neq 0$, which invalids the claim that $\bar\eta=2\eta$, and a whole region were dimensional reduction is broken with $\eta<\bar\eta< 2\eta$, as can be seen from the results in Fig. \ref{fig_etabareta_RFIM} obtained by the nonperturbative FRG of the equilibrium RFIM. The  description of the two regions clearly requires $3$ independent exponents. Since our work, large-scale computer simulations have confirmed the $3$-exponent scenario with $\bar\eta<2 \eta$ \cite{fytas_3d,fytas_4d}.

\section{Exact FRG and approximations}
\label{sec_NPFRG}

\subsection{Three possible formalisms for averaging over disorder}

There are several routes to carry out the average over the quenched disorder and derive functional RG flow equations for the cumulants of the renormalized disorder, which one can denote as:  ``Boltzmann-Gibbs", ``superfield", and ``dynamical". The first two specifically apply to the equilibrium behavior, the Boltzmann-Gibbs formalism being based on equilibrium partition function(al)s as in Eqs. (\ref{eq_partition_disorder}) and (\ref{eq_cumW}), the superfield one starting from the stochastic field equation in Eq. (\ref{eq_stochastic_GS}) and building a generating functional through the introduction of one additional auxiliary bosonic field and two auxiliary fermionic (Grassmannian) fields as put forward in Ref. [\onlinecite{parisi79}]. The third one can be used in equilibrium as well as in nonequilibrium situations and follows the Martin-Siggia-Rose-Janssen-Dedominicis \cite{MSR,janssen-dedom} construction of the generating functional based on the Langevin equation [Eq. (\ref{eq_stochastic_dynamics_RFIM})] in the Ito representation. 

We stress again that in all of the three formalisms one must introduce replicas or copies of the system, all with the same quenched disorder but coupled to {\it independent distinct sources}, which means that replica symmetry is explicitly broken. This is an unusual procedure in the cases of the superfield and the dynamical formalisms which are commonly considered as alternatives to the replica trick. The multi-replica procedure is necessary to provide a description of the cumulants with generic field arguments and therefore be able to account for the influence of avalanches and droplets on the long-distance behavior (see above).

The $3$ formalisms have benefits and drawbacks. The Boltzmann-Gibbs one is simpler but restricted to equilibrium and blind to the breaking or not of the underlying SUSY. The superfield construction is much more involved as in addition to dealing with superfields and superspace one needs to introduce an additional weighting of the solutions of Eq. (\ref{eq_stochastic_GS}) that generalizes the original Parisi-Sourlas construction \cite{parisi79} and allows one to recover ground-state dominance of the generating functionals at large scale. This is at the cost of dealing with a curved superspace. The upside is that the underlying supersymmetries of the theory can be explicitly incorporated and studied, with the derivation of associated Ward identities, and that the spontaneous breaking of SUSY, the rotation invariance in superspace, can be investigated. Finally, the main advantage of the dynamical formalism is that, especially in the case of the RFIM, the out-of-equilibrium behavior of the system when quasi-statically driven at zero temperature can be studied on an equal footing with the equilibrium behavior and the respective critical fixed points can be compared.  Furthermore, one can also describe the critical slowing down of the RFIM. 

A crucial point is that the exact FRG equations derived within the different formalisms of course coincide when applied to the same situation \cite{tissier06,tissier12,balog_activated,balog_eqnoneq}. In the following presentation we will consider the equilibrium behavior of in the presence of a random field and present the FRG in the context of the Boltzmann-Gibbs formalism which is conceptually simpler and requires lighter notations.

\subsection{Exact FRG equations for the cumulants}

The nonperturbative FRG is a version of Wilson's continuous RG \cite{wilson74,polchinski84,wetterich93} in which one progressively incorporates fluctuations of the local order parameter fields over larger length scales or shorter momenta. This can be done by introducing an ``infrared regulator'' that suppresses the integration over the modes with momentum $\vert q\vert$ less than some cutoff $k$ in the (functional) partition function and takes the form of a generalized ``mass'' (quadratic) term added to the bare action \cite{berges02}. Here and in most of what follows we present the formalism for the case of the RFIM ($N=1$) in equilibrium, which significantly alleviates the notations. The IR regulator added to the multi-copy action then reads
\begin{equation}
\begin{split}
\label{regulator}
\Delta S_k[\{\varphi_a\}]=
\frac{1}{2}\sum_{a,b}\int_{q}\varphi_a(q)  R_{k,ab}(q^2)\varphi_b(-q) \,,
\end{split}
\end{equation} 
where $\int_q\equiv \int d^{d} q/(2\pi)^d$ and $R_{k,ab}(q^2)=\widehat{R}_k\delta_{ab}(q^2) +\widetilde{R}_k(q^2)$. The functions $\widehat{R}_k(q^2)$ and $\widetilde{R}_k(q^2)$ are chosen to provide an infrared (IR) cutoff on all the fluctuations, which enforces a decoupling of the low- and high-momentum modes at the scale $k$. The function $\widehat{R}_k(q^2)$ adds a mass $\sim k^{2-\eta}$ to modes with $q^2<k^2$ and decays rapidly to zero for $q^2>k^2$, whereas the function $\widetilde{R}_k(q^2)$ (which must be chosen proportional to $-\partial_{q^2}\widehat{R}_k(q^2)$ to avoid an explicit SUSY breaking \cite{tissier11,tissier12}) reduces the fluctuations of the bare random field. 

Through this procedure, one defines the multi-copy generating functional of the correlation functions at scale $k$,
\begin{equation}
\begin{aligned}
\label{eq_Wk}
&W_k[\{J_a\}]=\\&
\ln \int \prod_a \mathcal D \varphi_a \, {\rm exp}\Big (- \sum_a S_B[\varphi_a] + 
\sum_a \int_x J_a(x)\varphi_a(x)  \\&
+ \frac 12 \sum_{a,b} \Delta_B \int_x  \varphi_a(x)\varphi_b(x) -  \Delta S_k[\{\varphi_a\}]\Big ) \,.
\end{aligned}
\end{equation} 

In the FRG approach, the central quantity is the ``effective average action'' $\Gamma_k$, the generating functional of the 1PI correlation functions at the scale $k$. It is obtained from $W_k[\{J_a\}]$ through a (modified) Legendre transform,
\begin{equation}
 \Gamma_k[\{\phi_a\}]+\Delta S_k[\{\phi_a\}]=-W_k[\{J_a\}]+\sum_a\int_x 
J_a(x)\phi_a(x),
\end{equation}
where the field $\phi_a=\delta W_k/\delta J_a(x)$ is the average of the physical field $\varphi_a$ in copy $a$. 

The evolution of the effective average action under the change of the IR cutoff $k$ is governed by an {\it exact} RG equation \cite{wetterich93},
\begin{equation}
\begin{aligned}
\label{eq_erge}
\partial_k\Gamma_k\left[\{ \phi_a\}\right ]= \frac{1}{2}\sum_{a,b}  \int_q  \partial_k R_{k,ab}(q^2) \big (\big[ \bm \Gamma _k^{(2)}+ \bm R_k\big]^{-1}\big)_{q,-  q}^{ab},
\end{aligned}
\end{equation}
where  $\bm \Gamma_k^{(2)}$ is the matrix formed by the second functional derivatives of $\Gamma_k$ with respect to the replica fields and the operator $\bm P_k[\{ \phi_a\}] \equiv [ \bm\Gamma _k^{(2)}+ \bm R_k]^{-1}$ is the exact propagator at the scale $k$. In physical terms, $\Gamma_k\left[\{ \phi_a\}\right ]$ is the (multi-replica) Gibbs free energy functional of the local order parameter fields obtained after a coarse-graining down to the momentum scale $k$. At the UV (or microscopic) scale $k= \Lambda$, $\Gamma_k$ essentially reduces to the bare replicated action, $\Gamma_\Lambda\approx \sum_a S_B[\varphi_a] -  (1/2) \sum_{a,b} \Delta_B \int_x  \varphi_a(x)\varphi_b(x)$ (for a Gaussian distributed bare random field),whereas at the end of the flow, when $k\to 0$, $\Gamma_k$ becomes equal to the full effective action (Gibbs free energy), $\Gamma_0=\Gamma[\{\phi_a\}]$.

Similarly to the full effective action $\Gamma[\{\phi_a\}]$ in Eq. (\ref{eq_cumg}),  $\Gamma_k[\{\phi_a\}]$ can be expanded in an increasing number of free replica sums,
\begin{equation}
\Gamma_k[\{\phi_a\}]=\sum_a 
\Gamma_{k1}[\phi_a]-\frac{1}{2}\sum_{a,b}\Gamma_{k2}[\phi_a,\phi_b]+\cdots \,,
\end{equation}
where $\Gamma_{k,p=1}$ is the disorder-averaged Gibbs free energy at scale $k$ and for $p\geq 2$ the $\Gamma_{kp}$'s are essentially the cumulants of the renormalized disorder at the scale $k$ \cite{tarjus04,tissier12}.

After expanding both sides of Eq. (\ref{eq_erge}) in an increasing number of free replica sums and using systematic algebraic manipulations, one obtains a hierarchy of exact RG flow equations for the cumulants of the renormalized disorder, $\partial_k  \Gamma_{k1}[\phi_a]=\cdots$, $\partial_k \Gamma_{k2}[ \phi_a , \phi_b]=\cdots$, etc., where the right-hand side of the flow equation for the $p$th cumulant retains the one-loop structure of Eq. (\ref{eq_propag_conn}) and involves up to the $(p+1)$th cumulant, so that all equations are coupled. The expressions also involve the exact ``connected" and ``disconnected" propagators, $\widehat{P}_k$ and $\widetilde{P}_k$, which are defined as
\begin{equation}
\label{1_copy_propk}
\widehat{P}_{k;x_1x_2}[\phi_a]=\big(\Gamma^{(2)}_{k1}[\phi_a]+\widehat{R}_k\big)^{-1}\big \vert_{x_1x_2}
\end{equation}
\begin{equation}
\begin{aligned}
\label{2_copy_propk}
\widetilde{P}_{k;x_1x_2}[\phi_a,\phi_b]=&-\int_{x_3x_4}\widehat{P}_{k;x_1x_3}[
\phi_a])\widehat{P}_{k;x_4x_2}[\phi_b] \times\\& 
\big (\Gamma^{(11)}_{k2;x_3x_4}[\phi_a,\phi_b]- \widetilde{R}_{k}(\vert x_3-x_4\vert\big ) \,,
\end{aligned}
\end{equation}
and which in the limit $k\to 0$ and for zero fields reduce to the physical connected and disconnected pair correlation functions of Eqs. (\ref{eq_propag_conn}) and (\ref{eq_propag_disc}).

\subsection{Nonperturbative approximation scheme}
\label{sec:NPapprox}

The hierarchy of exact FRG equations derived above cannot be solved exactly in general and we have proposed a systematic nonperturbative approximation scheme \cite{tarjus04,tissier06,tissier11,tissier12}. It consists in formulating an ansatz for the effective average action that relies on a joint truncation of (i) {\it the derivative expansion}, i.e., an expansion in the number of spatial derivatives for approximating the long-distance behavior of the 1PI correlation functions, and (ii) {\it the expansion in cumulants of the renormalized disorder}. (In the superfield/superspace formalism there is an additional truncation in increasing ``nonlocality in Grassmann space" \cite{tissier12} whereas in the dynamical formalism one also has to truncate the expansions in time derivatives and in powers of the response field \cite{balog_activated,balog_eqnoneq}.)

The choice of a minimal nonperturbative truncation is guided by a combination of factors: experience gained from studies on other models, constraints associated with the symmetries and supersymmetries of the theory, intuition or previous knowledge concerning the physics of the problem at hand, requirement of being able to recover as much as possible exact and perturbative results in the appropriate limits, and, of course, a practical limitation coming with the numerical ability to actually solve the set of FRG flow equations. For instance, it has been shown for many statistical mechanical models in the absence of quenched disorder, such as the $O(N)$ model, that a truncation of the derivative expansion at the second order gives a very good description of the asymptotic long-distance behavior \cite{berges02}. Furthermore, the convergence of the expansion has been found to be very rapid as one increases the order of the truncation \cite{wschebor19}. Concerning the truncation of the cumulant expansion, SUSY when it is present entails relations between the cumulant at order $p+1$ and the cumulant at order $p$, for any $p\geq 1$. The simplest nontrivial such relation (or Ward identity) implies for a uniform field that
\begin{equation}
\begin{split}
\label{eq_susy-ward_gamma2}
\Gamma_{k2}^{(11)}(q^2;\phi,\phi)\propto -\partial_{q^2}\Gamma_{k1}^{(2)}(q^2;\phi)\,.
\end{split}
\end{equation}
When SUSY is spontaneously broken, these Ward identities ceases of course to be satisfied. However, to avoid an {\it explicit} breaking of SUSY one must connect the order of the truncation of the derivative expansion to that of the cumulant expansion.

An efficient ansatz that can capture the long-distance physics including the influence of avalanches and droplets is then
\begin{equation}
\label{eq_ansatz_gamma1}
\Gamma_{k1}[\phi]=\int_x \big [U_k(\phi(x))+ \frac 12 
Z_k(\phi)(\partial_{x}\phi(x))^2 \big ]\,,
\end{equation}
\begin{equation}
\label{eq_ansatz_gamma2}
\Gamma_{k2}[\phi_1,\phi_2]=\int_x V_k(\phi_1(x),\phi_2(x))\,,
\end{equation}
and
\begin{equation}
\label{eq_ansatz_gammap}
\Gamma_{kp\geq3}=0 \,,
\end{equation} 
where the effective average potential $U_k(\phi)$ describes the thermodynamics of the system, $Z_k(\phi)$ is a function accounting for the renormalization of the field, and $V_k(\phi_1,\phi_2)$ is the $2$-replica effective average potential whose second derivative, $V_k^{(11)}(\phi_1,\phi_2)=\Delta_k(\phi_1,\phi_2)$,  is the second cumulant of the renormalized random field at zero momentum; $\Delta_k(\phi_1,\phi_2)$ is the key quantity that tracks avalanches and droplets through its functional dependence (see above). Inserting the above ansatz into the exact FRG equations for the cumulants leads to a set of coupled flow equations for the functions $U_k(\phi)$, $Z_k(\phi)$, and $V_k(\phi_1,\phi_2)$ [or, alternatively, the first derivative $U'_k(\phi)$, $Z_k(\phi)$, and $\Delta_k(\phi_1,\phi_2)$]. The RG is ``functional" as its central objects are functions instead of coupling functions and it is ``nonperturbative'' as the approximation scheme does not rely on an expansion in some small coupling constant or function.

Note that lower orders of the approximation scheme amount to taking $Z_k$ as a independent of the field $\phi$. According to Eq. (\ref{eq_susy-ward_gamma2}) this implies to consider $\Delta_k(\phi,\phi)$ also as a constant $\Delta_k$. The simplest implementation consists in assuming that $\Delta_k(\phi_1,\phi_2)=\Delta_k$ for all arguments, which, as we have argued in Sec. \ref{sec_avalanches}, prevents one from capturing the effect of avalanches and/or droplets. As a result such an approximation only predicts that the critical behavior is given by dimensional reduction \cite{tarjus_HRT}. On the other hand, the next higher order of the scheme consists in retaining terms up to O($\partial_x^4$) for the first cumulant, up to O($\partial_x^2$) for the second cumulant, and a nonzero but purely local third cumulant. This amounts to solving coupled partial differential equations for 5 functions of 1 field, 3 functions of 2 fields, and 1 function of 3 fields, a quite formidable numerical task which we have not yet achieved.

In the case of the $O(N)$ version of the random-field model, the analog of the truncation in Eqs. (\ref{eq_ansatz_gamma1}-\ref{eq_ansatz_gammap}) requires 2 field renormalization functions that depend on the variable $\rho=(1/2)\vert\bm\phi\vert^2$ and the $2$-replica potential is now a function of 3 fields, $\rho_1$, $\rho_2$, and $z_{12}=\bm\phi_1\mathbf{\cdot}\bm\phi_2/\sqrt{2\rho_1\rho_2}$. As a consequence of this increased numerical difficulty, we have also used some additional approximation in which we expand the dependence on $\rho_a$ of the functions around a nontrivial value of the field that minimizes the $1$-replica effective average potential (while keeping the full dependence in the other variable $z_{12}$) \cite{tissier06}.

\subsection{Dimensionless form of the nonperturbative FRG equations and fixed points}
\label{sec:dimensionless}

One more step is needed to cast the nonperturbative FRG flow equations in a form that is suitable for searching for the anticipated zero-temperature fixed points describing the critical behavior of the RFIM. One has to introduce appropriate scaling dimensions.  This requires defining a renormalized temperature $T_k$ which should flow to zero as $k \rightarrow 0$. (This is the precise meaning of a ``zero-temperature'' fixed point.) Near such a  fixed point, one has the following scaling dimensions: 
\begin{equation}
T_k \sim k^\theta,\;  Z_{k} \sim k^{-\eta}, \; \phi_a  \sim k^{\frac{1}{2}(d-4+\bar \eta)},
\end{equation}
with $\theta$ and $\bar \eta$ related through $\theta=2+\eta-\bar \eta$, and
\begin{equation}
U_k\sim k^{d-\theta}, \;V_k \sim k^{d-2\theta},
\end{equation}
so that the second cumulant of the renormalized random field $\Delta_k$ scales as $k^{-(2\eta- \bar \eta)}$.

Letting the dimensionless counterparts of $U_k, V_k,\Delta _k,  \phi$  be denoted by lower-case letters, $u_k, v_k,\delta _k,  \varphi$,  and expressing the results in terms of the dimensionless fields $\varphi=\frac{\varphi_1+\varphi_2}{2}$ and $\delta\varphi=\frac{\varphi_2-\varphi_1}{2}$, the resulting FRG flow equations can be symbolically written as
\begin{equation}
\begin{aligned}
\label{eq_flow_static}
&\partial_t u'_k(\varphi)=\beta_{u'0}(\varphi)+T_k\beta_{u'1}(\varphi)\,,\\&
\partial_t z_k(\varphi)=\beta_{z0}(\varphi)+T_k\beta_{z1}(\varphi)\,,\\&
\partial_t \delta_k(\varphi,\delta\varphi)=\beta_{\delta0}(\varphi,\delta\varphi)+T_k\beta_{\delta1}(\varphi,\delta\varphi)\,,
\end{aligned}
\end{equation} 
where $t=\log(k/\Lambda)$ is the dimensionless RG ``time" and a prime denotes a derivative for a function of a single argument. The ``beta functions", $\beta_{u'0}, \cdots, \beta_{\delta1}$, themselves depend on $u_k'$, $z_k$, $\delta_k$, their derivatives, and on the (dimensionless) cutoff functions defined from $\widehat R_k(q^2)=k^2Z_k \,\hat r(y=q^2/k^2)$, $\widetilde R_k(q^2)=\Delta_k\, \tilde r(y=q^2/k^2)=-\Delta_k \, \hat r'(y)$. In addition, the running anomalous dimensions $\eta_k$ and $\bar\eta_k$ are fixed by the conditions $z_k(0)=\delta_k(0,0)=1$ and reach fixed-point values when $k\to 0$ (or $t\to -\infty$). The expressions of the beta functions are given in Refs.~[\onlinecite{tissier06,tissier12,balog_activated}]. 

\begin{figure}[tb]
  \begin{center}
\includegraphics[width=\linewidth]{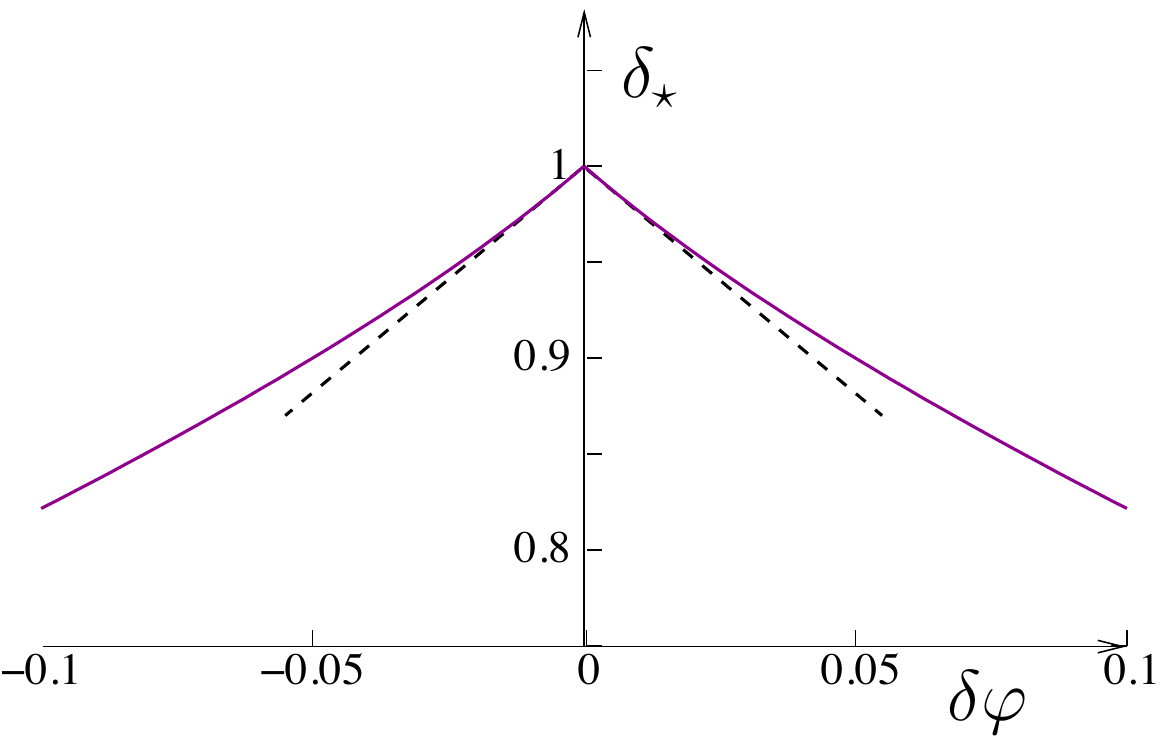}  
\caption{Dimensionless second cumulant of the renormalized random field $\delta_*(\varphi=0,\delta\varphi)$ at the equilibrium critical fixed point of the RFIM in $d=3$. One clearly sees the cusp in $\vert \delta\varphi \vert$ near the origin, i.e., when the replica fields become equal.}
\label{fig_cusp_delta}
\end{center}
\end{figure}

When $T=0$, the terms proportional to $T_k$ can be dropped in Eqs. (\ref{eq_flow_static}), and it is found that the fixed-point solution, which solves Eqs. (\ref{eq_flow_static}) with the left-hand side equal to zero (so that the renormalized theory displays scale invariance), displays two regimes: 

(1) for $d <d_{DR}\approx 5.1$, a ``cusp" in $\vert \delta\varphi \vert$ is present in the fixed-point function $\delta_*$ when $\delta\varphi\rightarrow 0$: 
\begin{equation}
\label{eq_cusp_delta}
\delta_*(\varphi, \delta \varphi) = \delta_{*}(\varphi,0) -\delta_{k,a}(\varphi)\vert \delta\varphi \vert + \frac{1}{2} \delta_{*,2}(\varphi)\delta\varphi^2  
+ O(\vert \delta \varphi \vert^3).
\end{equation}
This cusp, which is associated with the presence of avalanches on all scales at the critical point \cite{tarjus13}, is responsible for the breakdown of  dimensional reduction and SUSY \cite{tarjus04,tissier06,tissier11,tissier12}. It is illustrated in Fig. \ref{fig_cusp_delta} for the equilibrium RFIM in $d=3$. Note that this function (as the other fixed-point functions) is accessible by computer simulations through finite-size studies.

(2) For $d>d_{DR}$, the fixed-point function $\delta_*$ is ``cuspless", which ensures that the  dimensional-reduction property of the critical exponents, with, e.g., $\bar\eta(d)=\eta(d)=\eta_{{\rm ising},d-2}$ and $\theta=2$, is valid and that SUSY is obeyed at the fixed point. It is important to stress that avalanches are still present on all scales but only lead to a {\it subdominant cusp}:  $\delta_k(\varphi,\delta\varphi)=\delta_{k}(\varphi,0)-\delta_{k,a}(\varphi)\vert \delta\varphi \vert +O(\delta\varphi^2)$ where $\delta_{k,a}(\varphi)\sim k^\lambda$ when $k\rightarrow 0$ \cite{tarjus13}, with $\lambda >0$ characterizing the (diverging) number of spanning avalanches at criticality \cite{dahmen96,tarjus13}.

The passage from one regime to the other is very unusual as the cuspy fixed point emerges from the collapse of two cuspless fixed point in $d=d_{DR}$ through a mechanism of boundary layer \cite{baczyk_FP} (not to be confused with the ``thermal boundary layer" discussed just below).

When $T>0$, the cusp present in $\delta_k(\varphi,\delta\varphi)$ at $T=0$ is rounded due to temperature and $\delta_k$ develops a thermal boundary layer,
\begin{equation}
\label{eq_BL_ansatz}
 \delta_k(\varphi,\delta\varphi)=\delta_{k}(\varphi,0)+T_k b_k(\varphi,y=\frac{\delta\varphi^2}{T_k^2})+O(T_k^2,\delta\varphi^2)
\end{equation}
when $T_k,\delta\varphi \rightarrow 0$. One easily derives that the solution has the explicit form 
\begin{equation}
\label{eq_BL_solution}
b_k(\varphi,y)=\frac{a_{2*}(\varphi)}{a_{1*}(\varphi)}\Big(1-\sqrt{1+y\,\frac{a_{1*}(\varphi)^2\delta_{k,a}(\varphi)^2}{a_{2*}(\varphi)^2}}\Big),
\end{equation}
where $y\geq 0$, and the $a_{p*}$'s are (nonzero) fixed-point functions; $\delta_{k,a}(\varphi)$ behaves differently when $k\rightarrow 0$ for $d <d_{DR}$ and $d> d_{DR}$ (see above). Note that the boundary layer is the manifestation of a nonuniform convergence to zero temperature, i.e., to $T_k=0$, but the same zero-temperature fixed point is nonetheless reached whether one first takes $T=0$ or considers $T>0$. The key roles of the cusps in the functional dependence of the cumulants and of their rounding at finite temperature in a thermal boundary layer are found in the simpler case of an elastic manifold pinned in a random environment. There, the long-distance physics is accessible though a functional but perturbative RG in $d=4-\epsilon$, which allows for detailed analyses \cite{fisher86b,nattermann,narayan92,FRGledoussal-chauve,doussal_2loop,FRGledoussal-wiese,chauve_creep,balents-doussal_BL,ledoussal10} and provides some guidance for the more involved case of random-field models.

All of the above results can be generalized to the RF$O(N)$M, which in particular leads to the determination of the critical line $d_{DR}(N)$ and to the study of QLRO and its nontrivial  lower critical dimension $d_{lc}^{{\rm QLRO}}$ for $d\leq 4$ \cite{tissier06}: see also Sec. \ref{sec_summary}.

\section{Robustness of the nonperturbative results and comparison with perturbative analyses}
\label{sec_robustness}

\subsection{Why should one trust the outcome of the nonperturbative FRG?}

The above FRG approach is nonperturbative but approximate, which raises the question of the  robustness of its outcome.

We list below a number of arguments, not ranked by order of importance, which strengthen confidence in the results:

1) The nonperturbative FRG gives a consistent and unified description of the equilibrium behavior of random-field systems in the whole ($N$,$d$) diagram. The scenario of a critical line $d_{DR}(N)$ separating a region where dimensional reduction and SUSY are obeyed at the fixed point and a region where they are violated is in agreement with exact results, recent large-scale computer simulations, and perturbative FRG analyses in $d=4+\epsilon$ for the RF$O(N>1)$M and in $d=6-\epsilon$ at the two-loop level for the RFIM, as is discussed in more detail below. In the case of the RFIM the scenario is also supported by a recent loop expansion around the Bethe solution \cite{angelini19}.

2) The predicted critical exponents are in very good agreement with available computer simulation results in all dimensions. As shown in Fig. \ref{fig_etabareta_RFIM}, the anomalous dimensions $\eta$ and $\bar\eta$ are in good agreement with the best known values obtained by large-scale zero-temperature simulations. This is also true for the other critical exponents. (Unfortunately there are so far no computations of finite-size scaling functions in simulations, to which one could compare the nonperturbative FRG predictions for fixed-point functions.) Furthermore, the exponents satisfy all expected relations associated with scaling as well as the known rigorous bounds (e.g., $\eta\leq \bar\eta\leq 2\eta$ \cite{schwartz_exact}, $\eta\geq (4-d)/2$, $\bar\eta\geq 4-d$).

3) The nonperturbative FRG provides a description of the physics of random-field systems in terms of avalanches at zero temperature and droplets at finite temperature that is supported by real-space analyses in computer simulations \cite{vives_GS,wu-machta_GS,liu_GS,hartmann_droplet}. This description shares many similarities with the behavior of an elastic manifold in a random environment, for which many  FRG predictions have been successfully tested.

4) The minimal nonperturbative truncation of the FRG described in the preceding section is exact at one-loop level near the upper critical dimension $d_{uc}=6$ and near the lower critical dimension of the RF$O(N>1)$M for long-range ferromagnetism, $d_{lc}=4$. It is also exact in the large $N$ limit.

5) The nonperturbative FRG satisfies all the symmetries and supersymmetries of the theory and is able to describe their spontaneous breaking.

6) The approximation scheme is a systematic one and its quality can be tested. We have already checked the the robustness of the results on the choice of the IR cutoff functions \cite{tissier06,tissier12} and, for the RF$O(N)$M, on the additional approximations (field expansion) \cite{tissier06}. As for the accuracy of the truncation given in Eqs. (\ref{eq_ansatz_gamma1})-(\ref{eq_ansatz_gammap}) the best assessment would be to consider the next level of the approximation scheme. As mentioned before, this represents a very hard numerical task and we are still working on it. In the absence of such a test, one can nonetheless draw some conclusions. First, confidence in the truncation of the expansion in spatial derivatives comes from the evidence obtained from the study of simpler models without quenched disorder that the expansion is a powerful and rapidly converging method for describing the long-distance/small momentum sector of the theory \cite{berges02,wschebor19}. The truncation of the cumulant expansion (which is also constrained to that of the derivative expansion by the requirement of no explicit breaking of the underlying SUSY) is harder to assess. There is however an indirect way of doing it by considering its consequence on the exponent $\tau$ characterizing the distribution of avalanches sizes at criticality for the RFIM \cite{sethna93,dahman96,perkovic,perez04}. It can be shown, by following a procedure similar to that developed for an interface in a random environment \cite{doussal-wiese_aval}, that truncating at the order of the second cumulant of the renormalized random field as in Eq. (\ref{eq_ansatz_gamma2}) implies that the exponent $\tau$ is equal to $3/2$, its mean-field value, for all dimensions $d$ and whatever the level of truncation of the derivative expansion \cite{tarjus_unpublished}. It turns out, however, that this appears to be a rather good estimate of the exponent: In computer simulations, $\tau$ has been found to slightly increase with decreasing dimension, from $1.5$ in $d=6$ to $1.6$ in $d=2$ for the athermally and quasi-statically driven RFIM \cite{perkovic}, and from a recent careful study of the $3$-$d$ RFIM in equilibrium to be around $1.54$ \cite{liu09}. Overall, the error on the value of $\tau$ is thus of the order of $5\%$ or less. This suggests that the approximation neglecting the contribution of the third and higher cumulants is a reasonable one for describing the critical behavior of random-field systems \cite{footnote_universality}.

\subsection{Perturbative FRG analyses}

As mentioned above, an important property of the nonperturbative FRG is that because of its one-loop structure it reproduces through the minimal truncation the perturbative results obtained either near the lower critical dimension of long-range ferromagnetism for the RF$O(N>1)$M or near the upper critical dimension at the one-loop order. It is therefore important to check if the scenario remains valid when pushing the perturbative analyses to the two-loop order. This is what we now describe.

\subsubsection{Perturbative FRG for the RF$O(N>1)$M in $d=4+\epsilon$}

At the lower critical dimension for long-range ferromagnetic order for random-field models with continuous $O(N)$ symmetry, $d_{lc}=4$, the dimension of the fields is $(d-4+\bar\eta)/2=0$, so that the perturbative RG becomes {\it a priori} functional \cite{fisher85}. Fisher was the first to derive perturbative FRG equations for the RF$O(N)$M in $d=4+\epsilon$ at one loop \cite{fisher85}. After a first partial analysis given by Feldman \cite{feldman_RF}, we have provided a complete analysis of the one-loop perturbative FRG in Refs. [\onlinecite{tissier06,tissier_2loop,baczyk_FP}] and the results are as expected fully compatible with the nonperturbative FRG description. 

\begin{figure}[tb]
  \begin{center}
\includegraphics[width=\linewidth]{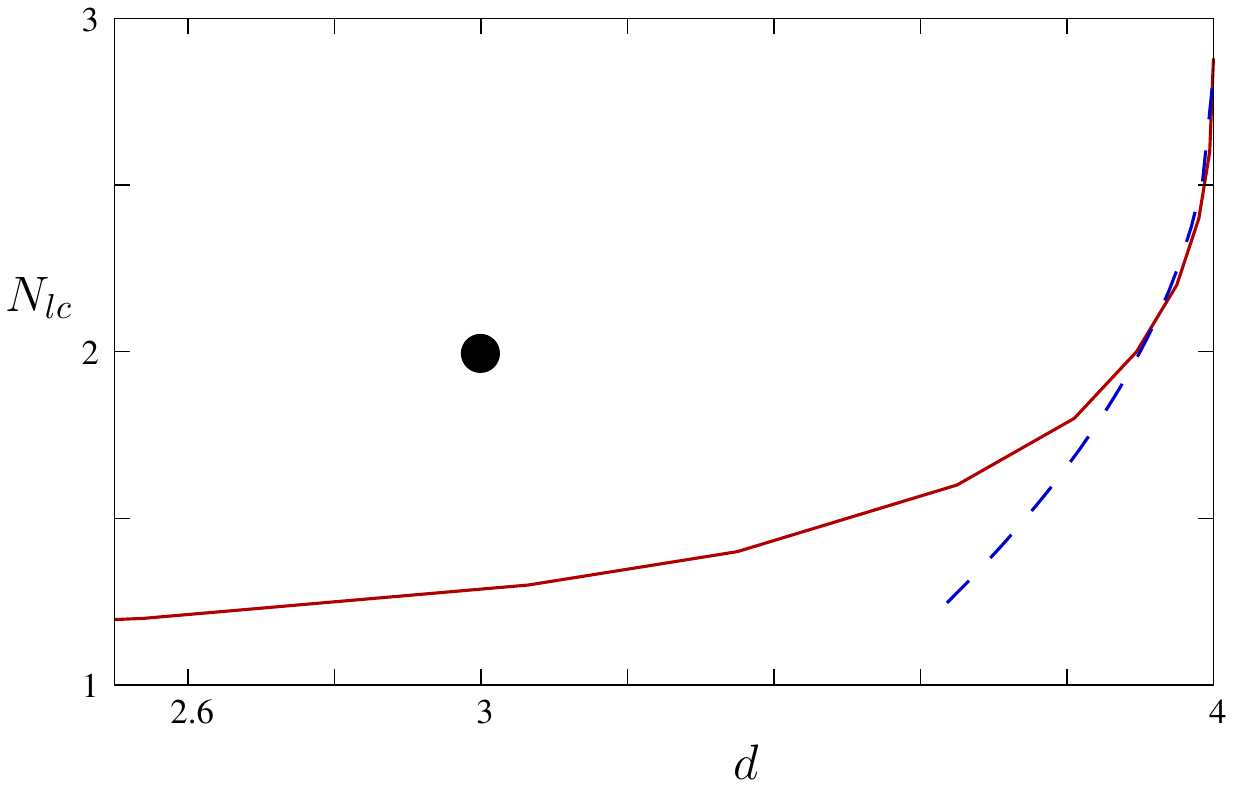}  
\caption{Lower critical dimension of QLRO for the RF$O(N)$M in equilibrium below $d= 4$: Comparison between the results of the two-loop perturbative FRG (dashed blue line) and of the nonperturbative FRG (full red line). The two curves start from $N_c=2.8347\cdots$ in $d=4$. The black circle denotes the physical case of the $XY$ model in $d= 3$, a case which is clearly below its lower critical dimension.}
\label{fig_QLRO}
\end{center}
\end{figure}

To go beyond this first step, one must consider the next order in the loop expansion. This can be done in a manner similar to that developed for the pure model at low temperature near $d= 2$, but with disorder now playing the role of temperature (temperature being irrelevant and eventually set to zero). The long-distance physics for weak disorder can be described in a field-theoretical setting by a {\it nonlinear sigma model}. The effective action is then perturbatively calculated in powers of the disorder correlator $R(z_{12})$. The latter is a function of a single variable $z_{12}$, which is the cosine between to replica fields that are unit vectors because of the fixed-length constraint. When going beyond the one-loop level, the technical difficulties become considerably more involved. On top of the rapidly increasing number of diagrams, diagrams which in the present case are functionals, the nonanalytic character of the renormalized effective action at $T= 0$ leads to the appearance of ``anomalous" terms in the diagrammatics, whose evaluation is {\it a priori ambiguous}. The solution of this problem (for a similar work, see Ref. [\onlinecite{doussal-wiese_RF}]) leads to a beta function for the renormalized disorder correlator (which is equivalent to $v_k(\rho_1,\rho_2,z_{12})$ in the above section with $\rho_1$ and $\rho_2$ sent to $\infty$). The analysis of the resulting fixed points fully confirm the one-loop and the nonperturbative FRG results concerning the restoration of dimensional reduction for a large enough value of $N$ (see Fig. \ref{fig_RFO(N)_phase_diagram}) when $d\geq 4$ ($\epsilon\geq 0$). In addition, it shows that the special point $d=4$, $N_c=2.8347\cdots$ does not correspond to a Berezinskii-Kosterlitz-Thouless transition at which the whole beta function would vanish in the loop expansion and that for $d\lesssim 4$ ($\epsilon\lesssim 0$)and $N\lesssim N_c$ a new, once unstable, fixed point appears, that describes the transition between paramagnetic and QLRO phases. This provides the mechanism by which the QLRO phase disappears below some critical dimension, in full agreement with the nonperturbative FRG predictions: see Fig. \ref{fig_QLRO}.

\subsubsection{Functional perturbation theory around the upper critical dimension for the RFIM}

Conventional perturbation theory, i.e., an expansion around a Gaussian reference theory in $\epsilon=6-d$, and the associated perturbative RG are known to fail  in low enough dimensions, as the $\epsilon$ expansion predicts at all orders dimensional reduction (see above). Since avalanches on all scales are always present at zero temperature and that breakdown of dimensional reduction is related to the existence of a singularity, a cusp, in the functional dependence of the renormalized cumulants of the random field {\it at the fixed point}, perturbation theory can still be useful provided one upgrades it to a functional approach. So long as the cuspless fixed point exists below $d=6$, one can indeed take into account the effect of the avalanches. This effect is then subdominant near the fixed point and is characterized by an exponent $\lambda>0$, which also allows one to compute the fractal dimension $d_f$ of the largest typical avalanches at criticality (see above).

This new twist in the perturbation theory is made possible by studying \textit{cuspy perturbations}, i.e., by including in the theory functional operators that are nonanalytic in the field dependence \cite{tarjus_perturb}. This can be done, e.g., in the RFIM, by (i) adding to the bare action an ``anomalous" contribution 
\begin{equation}
  \label{eq:anom-lag}
S_{\text{cusp}}=\frac {w_B}{4 }\int_x \sum_{a,b=1}^n \varphi_{B,a}(x) \varphi_{B,b}(x) |\varphi_{B,a}(x) -\varphi_{B,b}(x)|\,,
\end{equation}
where only the operator with the lowest canonical dimension is considered, and (ii) perturbatively renormalizing the amplitude of this contribution through a loop expansion \cite{tarjus_perturb}. The lowest nonzero correction term appears at the two-loop level. It leads for instance to a determination of the exponent $\lambda$ in powers of $\epsilon$:
\begin{equation}
\label{eq_lambda_eps}
\lambda=1- \frac 12 \epsilon -\frac 5{36}\epsilon^2 +\mathcal O(\epsilon^3).
\end{equation}
Clearly, if $\lambda$ becomes negative, perturbation theory breaks done (it can also break down before this happens). The above result indeed shows that $\lambda$ decreases as $d$ decreases below $6$ and indicates that $\lambda$ would go to $0$ at some nontrivial point \cite{tarjus_perturb}. This behavior is a confirmation of the predictions of the nonperturbative FRG \cite{tarjus13,baczyk_FP}.  The fractal dimension $d_f$ predicted from Eqs. (\ref{eq_lambda_df}) and (\ref{eq_lambda_eps}) is shown in Fig. \ref{fig_df}.

Note finally that the fact that singular corrections are present but still subdominant near the upper critical dimension is also supported by a recent perturbative loop expansion around the Bethe solution \cite{angelini19}.

\section{Further results}
\label{sec_further}

The nonperturbative FRG approach has also been extended to study more more systems and phenomena in the presence of random fields, which we briefly summarize below.

\subsubsection{Equilibrium critical behavior of the RFIM with long-range interactions and long-range disorder correlations \cite{baczyk_3DLR,balog_1DLR}} 

Aside from relevance to physical systems, the interest in long-range models comes from the fact that the presence of long-range interactions, which are power-law decaying with distance as $r^{-(d+\sigma)}$, decreases the lower critical dimension of a model and that varying the exponent $\sigma$ of the power law in a fixed dimension $d$ allows one to find a spectrum of critical behavior that goes from mean-field for truly long-range  interactions ($\sigma\leq \sigma_{uc}$) to the absence of transition for short-range decay ($\sigma\geq \sigma_{lc}$) while spanning a continuous range of nonclassical behavior in between. In some sense, changing the exponent $\sigma$ at fixed $d$ is like changing the dimension $d$ in a short-range model. In the case of the RFIM, this has the merit to bring the critical passage at $\sigma_{DR}(d)$ from a regime where critical behavior is dominated by avalanches to a regime where avalanches only play a subdominant role to physically accessible dimensions, $d\leq 3$. Actually, in $d=1$ (and $d=2$ as well) there can be no proper $d\to d-2$ dimensional reduction but a critical value of $\sigma$ nonetheless separates a region where the fixed point is  ``cuspy" from a region where the fixed point is ``cuspless" \cite{balog_1DLR}. In $d=3$ on the other hand, the breaking of SUSY and the associated dimensional-reduction breakdown can be investigated by also considering long-range correlations of the bare disorder, with a power-law decay of the correlator $\Delta_B(r)\sim r^{-(d-\rho)}$  with distance $r$ \cite{bray_LR,baczyk_3DLR}. SUSY then requires that $\rho=2-\sigma$, but it is violated, as is dimensional reduction, below some $\sigma_{DR}\approx 0.72$, which is intermediate between $\sigma_{lc}=1$ and $\sigma_{uc}=1/2$. To obtain all the results, the formalism sketched in previous sections has to be extended to include the singular momentum dependence of the vertices resulting from the long-range nature of the interactions and of the disorder correlations \cite{baczyk_3DLR,balog_1DLR}.

\subsubsection{Activated dynamic scaling for the critical slowing down of the RFIM \cite{balog_activated}}

By upgrading the nonperturbative FRG of the RFIM to the dynamical formalism, one can study the critical slowing down in the relaxation to equilibrium near the critical point. Activated dynamic scaling, in the form described in Sec. \ref{sec_recap}, is then obtained for all dimensions $d<d_{uc}=6$. The physical reason behind this dynamic scaling is the presence of power-law rare droplets at low temperature \cite{fisher_activated}. This is captured in the FRG through the thermal rounding of the cusp in the cumulants of the renormalized random field and the dangerous irrelevance of the temperature. The exponent $\psi$ characterizing the divergence of the effective activation barrier [see eq. (\ref{eq_activated_scaling})] is predicted to be equal to the temperature exponent, $\psi=\theta$, for $d\leq d_{DR}\approx 5.1$ and to decrease as $\psi=\theta -2\lambda$ where $\lambda$ is the exponent characterizing the irrelevance of the avalanches near the zero-temperature fixed point, for $d\geq d_{DR}$ \cite{balog_activated}. Above the upper critical dimension $d_{uc}=6$ at which $\theta=2$ and $\lambda=1$, activated dynamic scaling gives way to conventional critical slowing down (with the dynamical exponent $z=2$).

\subsubsection{Criticality in the RFIM in and out of equilibrium \cite{balog_eqnoneq}}

As explained in Secs. \ref{sec_models} and \ref{sec_recap}, the RFIM can be studied in equilibrium but also out of equilibrium, where it displays a phase transition as a function of disorder strength when it is quasi-statically driven by an applied source at zero temperature. The process leads to hysteresis and the out-of-equilibrium critical points found along the hysteresis branches come with scale-free ``dynamic" avalanches (crackling noise). Quite strikingly, despite the fact that one type of critical point is at equilibrium and at zero external field while the other is out of equilibrium and at a nonzero value of the applied external field, and that they take place at different values of the disorder strength, the two critical behaviors are characterized by exponents that have been found very close in numerical simulations \cite{maritan94,perez04,liu09}. The nonperturbative FRG in the dynamical formalism allows one to treat the two situations on an equal footing and to derive functional flow equations that properly describe the two different protocols (as first implemented in a perturbative context in the FRG of a random elastic manifold in the equilibrium pinned phase and near the depinning transition \cite{doussal-wiese}). It has been shown that in spite of the similarity of the critical exponents and of some scaling functions, the two critical behaviors are {\it not} in the same universality class and are controlled by distinct zero-temperature fixed points whenever $d<d_{DR}\approx 5.1$ \cite{balog_eqnoneq}. The signature of this difference is more easily detected by looking at the $Z_2$ or $Z_2$-broken shape of some fixed-point functions. Above $d_{DR}$ on the other hand, both types of critical points are controlled by the ``cuspless" dimensional-reduction fixed point.

\subsubsection{Higher-order random anisotropies in $O(N)$ models in equilibrium \cite{tissier06,tissier_2loop,mouhanna_RA}}

When the theory has an underlying continuous $O(N)$ symmetry, quenched disorder can take the form of random anisotropies that couple to products of field components. If these random anisotropies are only of even ranks, the model has an additional inversion symmetry compared to the random field model studied above in this article. The starting point of the theoretical description is a bare action similar to that in Eq. (\ref{eq_ham_dis_RFIM}) but with $S[\bm\varphi; \bm h]=  S_B[\bm\varphi]-\int_{x} \sum_{\mu,\nu=1}^N \tau^{\mu\nu}(x)\varphi^\mu(x)\varphi^\nu(x)$ with the random anisotropy tensor $\bm \tau$ sampled from a Gaussian distribution with zero mean and variance $\overline{\tau^{\mu\nu}(x)\tau^{\rho\sigma}(y)}=(\Delta_2/2)(\delta_{\mu\rho}\delta_{\nu\sigma}+\delta_{\mu\sigma}\delta_{\nu\rho})\delta^{(d)}(x-y)$. Such a random anisotropy $O(N)$ model [RA$O(N)$M] with $N=2$ ($XY$) and $N=3$ (Heisenberg) describes the critical physics of amorphous magnets such as rare-earth compounds \cite{harris73} and of nematic liquid crystals in a disordered porous medium \cite{feldman_nematic}. The same FRG treatment developed for the RF$O(N)$M applies here, i.e., both an approximate nonperturbative method \cite{tissier06} and a perturbative analysis up to two loops near $d=4$ \cite{tissier_2loop}. It allows for a full description of dimensional-reduction breaking and of QLRO. Interestingly, the RA$O(N)$M has also a nontrivial behavior with putative ``glassy" phases in the large $N$ limit, and this is accessible through an FRG treatment \cite{mouhanna_RA}.

\section{Conclusion}
\label{sec_conclusion}

The functional renormalization group, in its nonperturbative implementation complemented when possible by perturbative analyses near the upper or the lower critical dimension, provides a complete theoretical description of the long-distance (and long-time) physics of the random-field Ising and $O(N)$ models. As such it has helped to solve most of the pending puzzles concerning random-field systems. The strength of the  approach are (i) a unified account of the whole domain of $N$, which can be continuously varied from $1$ (Ising) to $\infty$, and $d$, which can be continuously varied from the lower to the upper critical dimension; (ii) a description of singular collective events, such as avalanches present at zero temperature and droplets present at low temperature, through proper functional dependences of the renormalized disorder cumulants; (iii) a nonperturbative treatment which, e.g., gives access, even away from any perturbative regime, to the nontrivial critical dimension $d_{DR}(N)$ below which dimensional reduction and SUSY break down; (iv) predictions for the critical exponents (and scaling functions) that are in very good agreement with computer simulation results, when available, and that satisfy all expected relations associated with scaling and known exact bounds; (v) an easy implementation of all symmetries and supersymmetries of the theory, as well as a framework to study their possible spontaneous breaking, (vi) the formulation of a systematic approximation scheme.

The nonperturbative FRG approach is also a versatile method that can be applied to the study of other disordered systems. This is for instance readily done for models with with a random mass \cite{tarjus08}, random anisotropies \cite{tissier06,mouhanna_RA}, for a disordered Bose fluid \cite{dupuis19}, and for an elastic manifold in a random environment \cite{balog_REMM}. More generally, this can be carried out for any disordered model whose local order parameter is a simple field as magnetization, density or displacement. The method can also be used to investigate nonuniversal quantities: Starting from a microscopic model defined on a lattice, one can generalize the lattice nonperturbative RG of Refs. [\onlinecite{dupuis_lattice}] to compute critical temperature and phase diagram, actual length scales, etc. On the other hand, extension to the problem of spin glasses is harder due to the composite nature of the local order parameter \cite{book_SG}, which is an overlap between two configurations of the system, and remains, yet, to be satisfactorily implemented.
\\

%To be done?????:

%- One more step in the truncation scheme to pinpoint the value of $d_{DR}$. Also a BMW scheme with either an enforcement of SUSY for estimating $d_{DR}$ or not when SUSY is strongly broken as in $d=3$.

%- For the RFIM: Exponent $\tau$ of the power-law distribution of avalanche sizes at criticality. With the present truncation at the second cumulant at the level of the 1PI generating functionals one finds $\tau=3/2$, which is the mean-field result \cite{tau-mean-field}, in all dimensions $d$ (tarjus, unpublished). This offers an indirect way to assess the validity of the truncation. The exponent measured in simulations at zero temperature either in or out of equilibrium is always close to $1.5$ up to ????.

%- To obtain nonuniversal quantities in specific systems: need the lattice nonperturbative RG, e.g., to study the difference (or not) between RFIM and dilute antiferromagnet in a uniform field. At least including the third cumulant in the next order of the truncation scheme would allow to check universality of the long-distance physics with respect to the bare distribution of the random field. Would a zero and a nonzero initial value of the third cumulant lead to different fixed points?

\begin{acknowledgements}
We thank Ivan Balog (Institute of Physics, Zagreb, Croatia) with whom most of the recent work on the nonperturbative FRG of the RFIM has been performed.
\end{acknowledgements}


\begin{thebibliography}{85}
\bibitem{footnote_arguments}
Arguments here refer to replica fields, not to spatial coordinates or momenta, as discussed in Sec. \ref{sec_avalanches}.
\bibitem{footnote_universality}
This conclusion is also supported by recent large-scale zero-temperature simulations of the RFIM in $d=3$ \cite{fytas_3d} that gave very strong evidence that the critical behavior is independent of the distribution of the random field. Cumulants of the (bare) disorder of higher order than  $2$ do not modify the universal properties.
\end{thebibliography}
\end{document}